\documentclass[a4paper,11pt]{article}
\usepackage{jheppub}
\usepackage[latin1]{inputenc} 
\usepackage[T1]{fontenc}
\usepackage[safe]{textcomp}
\usepackage{lmodern} 
\usepackage{epsfig}
\usepackage{float}
\usepackage{amstext}  
\usepackage{amsfonts}
\usepackage{amsthm}
\usepackage{bm}       
\usepackage[bottom]{footmisc} 
\usepackage{array}                
\usepackage{xcolor} 
\usepackage{framed}
\usepackage{slashed}
\usepackage[absolute]{textpos}
\usepackage{axodraw4j}
\usepackage{dsfont}
\usepackage{multirow}

\newcommand{\p}{\partial}

\newcommand{\f}[2]{\frac{#1}{#2}}
\newcommand{\sss}[1]{\scriptscriptstyle#1}

\newcommand{\bea}{\begin{eqnarray}}
\newcommand{\eea}{\end{eqnarray}}
\newcommand{\be}{\begin{equation}}
\newcommand{\ee}{\end{equation}}
\newcommand{\ba}{\begin{align}}
\newcommand{\ea}{\end{align}}
\newcommand{\beas}{\begin{eqnarray*}}
\newcommand{\eeas}{\end{eqnarray*}}
\newcommand{\bes}{\begin{equation*}}
\newcommand{\ees}{\end{equation*}}
\newcommand{\bas}{\begin{align*}}
\newcommand{\eas}{\end{align*}}
 \newcommand{\rig}{\rightarrow}
 
\newcommand{\ssL}{{\mathcal L}} 
\newcommand{\eps}{{\varepsilon}}
\newcommand{\cd}{{\cdot}} 
\newcommand{\cf}{C_{\scriptscriptstyle{F}}} 
\newcommand{\ca}{C_{\scriptscriptstyle{A}}}
\newcommand{\tr}{T_{\scriptscriptstyle{F}}}

\newcommand{\dR}{d_{\scriptscriptstyle{R}}}
\newcommand{\Ng}{n_{\scriptscriptstyle{g}}}

\newcommand{\Nc}{N_{\scriptscriptstyle{c}}}
\newcommand{\Nf}{n_{\scriptscriptstyle{f}}}
\newcommand{\gs}{g_{\scriptscriptstyle{s}}}

\newcommand{\als}{\alpha_{\scriptscriptstyle{s}}}
\newcommand{\as}{a_{\scriptscriptstyle{s}}}
\newcommand{\lb}{\left(}
\newcommand{\rb}{\right)}

\definecolor{bluemar}{rgb}{0,0,.5}
\definecolor{redmar}{rgb}{.8,0,0}
\definecolor{greenmar}{rgb}{0,.5,0}

\allowdisplaybreaks
\def\bbuildrel#1_#2^#3%
{\mathrel{\mathop{\kern 0pt#1}\limits_{#2}^{#3}}}

\newcommand{\ice}[1]{\relax}
\newcommand{\beq}{\begin{equation}}
\newcommand{\eeq}{\end{equation}}
\newcommand{\re}[1]{(\ref{#1})}

\title{OPE of the pseudoscalar gluonium correlator in massless QCD to three-loop order}
\author[a]{M. F. Zoller}
\affiliation[a]{Institut f\"ur Theoretische Teilchenphysik, Karlsruhe
  Institute of Technology (KIT), \mbox{D-76128 Karlsruhe, Germany}}

\emailAdd{max.zoller@kit.edu}

\abstract{In this paper analytical results are presented for higher order corrections to coefficient
functions of the operator product expansion (OPE) for the
correlator of two pseudoscalar gluonium operators $\tilde{O}_1=G^{\mu \nu}\tilde{G}_{\mu \nu}$.
The Wilson coefficient in front of the scalar gluon condensate operator $O_1=-\f{1}{4} G^{\mu \nu}G_{\mu \nu}$ is
given at three-loop accuracy.
The leading coefficient $C_0$ in front of the unity operator $O_0=\mathds{1}$ 
has been calculated up to three-loop order some time ago \cite{Chetyrkin:1998mw} but has been checked independently in this work.
It is interesting to see that the coefficient $C_1$ in the pseudoscalar case is finite, 
whereas contact terms appear in $C_0$ in this case and in both coefficients $C_0$ and $C_1$
in the cases of the scalar gluonium correlator and the energy momentum tensor
correlator \cite{Zoller:2012qv}.
For the corresponding Renormalization Group invariant Wilson coefficients which are also constructed
the results are partially extended to four-loop accuracy.
All results are given in the $\overline{\text{MS}}$-scheme at zero temperature.
}

\subheader{TTP13-003\\SFB/CPP-13-04}

\begin{document}
\maketitle

\setlength{\fboxrule}{0.5 mm} 

\section{Motivation}
Euclidian correlators of local operators are important objects in quantum field theory.
Firstly, they have many important applications, e.g. in sum rules, where they are
connected to physical quantities like spectral densities through dispersion relations.
Secondly, they often have interesting properties in themselves, like their non-trivial renormalization,
which are important for the understanding of quantum field theories.
Such correlators are defined in momentum space as             
\be 
i\int\!\mathrm{d}^4x\,e^{iqx} T\{\,[O](x)[O](0)\}
\label{corrOP}
\ee
with a large Euclidian momentum $q$. Here and in the following the squared brackets indicate
that the renormalized form of some operator $O$ is used.
Usually, we are interested in the vacuum expectation value (VEV) of the
correlator
\be 
\Pi(Q^2)=i\int\!\mathrm{d}^4x\,e^{iqx}\,\langle 0|T\{\,[O](x)[O](0)]\}|0 \rangle \qquad (Q^2=-q^2) 
\label{corrVEV}
\ee
which can be calculated in perturbation theory. But if we take $|0 \rangle$ to be the physical vacuum state
we also have to consider non-perturbative effects. Starting from the perturbative region of momentum space this is done
by means of an operator product expansion (OPE). The idea is to expand the bilocal operator product \re{corrOP} in a series
of local operators with Wilson coefficients depending on the large Euclidean momentum q \cite{wilson_ope}:
\footnote{Effectively this expansion separates the high energy physics, which is
contained in the Wilson coefficients, from the low energy physics which
is taken into account by the VEVs of the local operators, the
so-called condensates \cite{Shifman:1978bx}. These cannot be calculated in perturbation
theory, but need to be derived from low energy theorems or be
calculated on the lattice.}
\bea 
i\int\!\mathrm{d}^4x\,e^{iqx} T\{\,[O](x)[O](0)\}&=&\sum \limits_i C_{i}^B(q) (Q^2)^\f{2 \text{ dim}(O)-\text{dim}(O_{i})-4}{2} O_{i}^B
\label{OPE_B}\\
&=&\sum \limits_i C_{i}(q) (Q^2)^\f{2 \text{ dim}(O)-\text{dim}(O_{i})-4}{2} [O_{i}]{},
\label{OPE_R}
\eea
where the index $B$ marks bare quantities and the factor $(Q^2)^\f{2 \text{ dim}(O)-\text{dim}(O_{i})-4}{2}$ 
constructed from the mass dimensions of the operators involved makes
the Wilson coefficients $C_i(q)$ dimensionless.

In a sum rule approach to glueballs three operators are usually investigated as insertions
on the lhs of \re{OPE_B} (see e.g. \cite{forkel_sumrule}):
 \begin{align}
     O_1(x)  &=-\f{1}{4} G^{a\,\mu \nu}G^a_{\mu \nu}(x)  & \text{(scalar)} {},\label{O1}\\
     \tilde{O}_1(x)  &=G^{a\,\mu \nu}\tilde{G}^a_{\mu \nu}(x)  & \text{(pseudoscalar)}{},\label{O1t}\\ 
     O_T^{\mu \nu}(x) &=T^{\mu \nu}(x) \label{OT} & \text{(tensor)}{},
    \end{align} 
where $G^a_{\mu \nu}=\partial_\mu A^a_\nu-\partial_\nu A^a_\mu+g f^{abc}A^b_\mu A^c_\nu$ is the gluon field strength tensor,
\be
\tilde{G}^a_{\mu \nu}=\eps_{\mu\nu\rho\sigma}G^{a\,\rho \sigma}
\ee
the dual gluon field strength tensor and $T^{\mu \nu}$ the energy-momentum tensor of QCD.
Having discussed the correlators of $O_1$ and $O_T^{\mu \nu}$ in \cite{Zoller:2012qv}
the results for the correlator of \re{O1t}
\be 
X_t(q):= i\int\!\mathrm{d}^4x\,e^{iqx}T\{\, [\tilde{O}_1](x)[\tilde{O}_1](0)\},
\label{corrO1s}
\ee
whose VEV $\chi_t(q):=\langle 0|X_t(q)|0\rangle$ is also known as the topological susceptibility of
QCD\footnote{For a discussion of topological effects
in QCD and the significance of the operator $\tilde{O}_1$ and the correlator \re{corrO1s} 
in that respect see e.g. \cite{Luscher:2004fu, PhysRevLett.94.032003}.}, are presented here. 
This correlator has been connected to the mass of the $\eta^{'}$-meson through the Witten-Veneziano formula 
\cite{Witten:1979vv,Veneziano:1979ec,Seiler:2001je,Giusti:2001xh}:
\be
\left.\f{\als^2}{32 i \pi^2}\chi_t(q)\right|_{q\rightarrow 0, \f{\Nf}{\Nc}\rightarrow 0}
=\f{m_{\eta^{'}}^2 F_\pi^2}{\Nf}\quad \text{(leading order)}{},
\ee
where $F_\pi\approx 94$ MeV is the pion decay constant. An explicit sum rule calculation with an OPE at one-loop level using a Borel
transformation has been done in \cite{Novikov:1979ux}. In this work the value $m_{\eta^{'}}\approx 1$ GeV is correctly estimated.
A similar analysis at two-loop level but using only the leading coefficient $C_0$ has been done in \cite{Kataev:1981aw}.\footnote{It will be shown however
in section \ref{chapNumerics} that the $\als$-expansion of the Wilson-coefficients, especially of $C_0$ converges rather badly at the low scales
considered in these analyses. This should be taken into account in the treatment of pseudoscalar hadrons within the sum rule approach.}

The correlator defined in \re{corrOP} with renormalized operators is finite, i.e. all its matrix elements are finite,
except for possible contact terms. These arise from the point where \mbox{$x\equiv 0$} and manifest themselves as divergences
$\propto \delta(x)$ and derivatives of $\delta(x)$ or in momentum space terms polynomial in $q$. These local terms
do not contribute to sum rules and can and should be subtracted with proper counterterms.

The leading term on the rhs of \re{OPE_B} is the coefficient in front of the unit operator $\mathds{1}$
which is just the perturbative VEV of the correlator \re{corrOP}:
\be
(Q^2)^2 C_0(q)=\langle 0|X_t(q)|0\rangle|_{\text{pert}}.
\ee
The coefficient $C_0$ is known for the scalar case \re{O1} at four-loop level \cite{Baikov:2006ch} and for the pseudoscalar
case \re{O1t} \cite{Chetyrkin:1998mw} and the energy-momentum tensor correlator \cite{Zoller:2012qv} at three-loop level.\footnote{Two-loop results for $C_0$ in the scalar and pseudoscalar case \cite{Kataev:1982gr} and in the
tensor case in gluodynamics ($\Nf=0$) \cite{Pivovarov_tensorcurrents} have been known for a long time.}
The next important contribution in the OPE is the coefficient of the dimension four operator $[O_1]$ \re{O1}.\footnote{In the
case of massive fermion flavours $f$ we would also have contributions proportional to the dimension two operator $O^f=m_f^2\,
\mathds{1}$ and the dimension four operator $O_2^f=m_f\bar{\psi_f}\psi_f$. In the case of temperature $T\neq 0$ Lorentz variant operators like $T^0_0 \sim e+p$ 
with the energy density $e$ and the pressure $p$ have to be considered as well. At $T=0$, however, only Lorentz and gauge invariant
scalar operators contribute to the the VEV in \re{corrVEV} which is the quantity that we are ultimately interested in.
For a discussion of the correlator $X_t(q)$ at finite temperature up to $\mathcal{O}(\als)$ see \cite{Laine:2010tc}.}
The coefficient $C_1$ has been calculated at two-loop level for the scalar\footnote{The one-loop result for the scalar case was first derived
in \cite{Novikov_scalargluonium}.} and tensor cases \cite{Zoller:2012qv}. 
Here we present the coefficent $C_1$ for the pseudoscalar case at three-loop level
which so far has only been known to one-loop accuracy \cite{Novikov:1979ux, Zhang:2003mr}.

All physical matrix elements of $[O_1]=Z_G O_1^B$ are finite and so is the renormalized 
coefficient $C_1$:\footnote{In the massless case $O_1$ only mixes with unphysical operators whose matrix elements with physical external
states vanish. The renormalization of $O_1$ including these unphysical contributions as well 
as the mixing with $O_2^f$ in the massive case can be found in \cite{Spiridonov:1984br}.}
\be C_1=\f{1}{Z_G}C_1^B. \label{C1ren}
\ee
The renormalization constant
\be Z_G=1+\als\f{\p}{\p\als}\ln Z_{\als}=\lb 1- \f{\beta(\als)}{\eps}\rb^{-1} \ee
has been derived in a simple way in \cite{Spiridonov:1984br} (see also an earlier work
\cite{Nielsen:1975ph}).
Here $Z_{\als}$ is the
renormalization constant\footnote{Often in the
literature $Z_{\als}$ is used instead of $Z_G$ and $\als G^{\mu
\nu}G_{\mu \nu}$ instead of $O_1$. This renormalization is only valid up to
first order in $\als$ as the renormalization constants $Z_G$ and
$Z_{\als}$ coincide to this accuracy.  In higher orders, however, $Z_G$ and $Z_{\als}$
differ.} for $\als$ and the $\beta$-function is defined as
\beq\beta(\als)=
\mu^2\f{\mathrm{d}}{\mathrm{d}\mu^2}\, \ln \als
=  - \sum_{i \ge 0} \beta_i \, \left( \frac{\als}{\pi} \right)^{i+1}
{}.
\label{be:def}
\eeq
The outline of this paper is as follows. In the next section the renormalization properties of $\tilde{O}_1$ 
will be discussed. In section \ref{chapCalcandRes} the details of the calculation will be described (section \ref{chapSetup}) and the results for the 
OPE of \re{corrO1s} will be presented (section \ref{chapResults}). After that Renormalization Group invariant (RGI) operators and Wilson coefficients will be 
constructed (section \ref{chapRGIsection}) followed by a numerical evaluation of the main results (section \ref{chapNumerics}).
Finally, some conclusions and acknowledgements will be given.

\section{Renormalization of $\tilde{O}_1$ and its correlator} \label{chapRen}
The operator $\tilde{O}_1$ forms a closed set under renormalization with the pseudoscalar fermionic operator
\be \partial_\mu J_5^\mu:= \eps^{\mu\mu_1\mu_1\mu_3} 
\partial_\mu \sum \limits_f \bar{\Psi}_f \gamma_{\mu_1}\gamma_{\mu_2}\gamma_{\mu_3} \Psi {},\label{dJ5Larin} \ee
which can be written as
\be \partial_\mu J_5^\mu= \partial_\mu \sum \limits_f \bar{\Psi}_f \gamma^\mu \gamma_5 \Psi \label{dJ5} \ee
in the Larin scheme for $\gamma_5$ \cite{Larin:1993tq}.

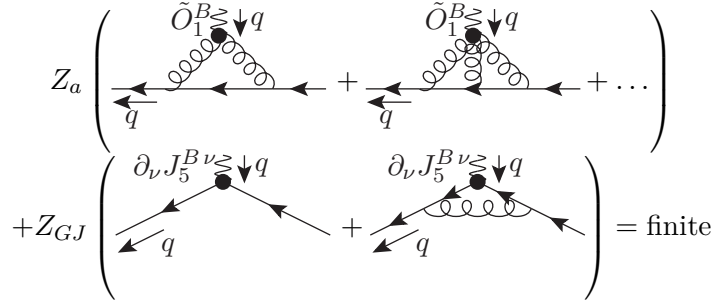
\begin{figure}[h!]
\begin{center}
$Z_{a} \left(
  \begin{picture}(80,30) (0,0)
    \SetWidth{0.5}
    \SetColor{Black}
\Photon(40,30)(40,20){3}{2.5}
\ArrowLine(80,0)(60,0)
\ArrowLine(60,0)(20,0)
\ArrowLine(20,0)(0,0)
\Gluon(40,20)(20,0){3}{4.5}
\Gluon(40,20)(60,0){3}{4.5}
\Vertex(40,20){3}
\LongArrow(48,30)(48,23)
\LongArrow(17,-5)(4,-5)
    \Text(38,20)[rb]{{\Black{$\tilde{O}_1^B$}}}
    \Text(52,22)[lb]{{\Black{$q$}}}
    \Text(5,-15)[lb]{{\Black{$q$}}}
  \end{picture}
\,+
  \begin{picture}(80,30) (0,0)
    \SetWidth{0.5}
    \SetColor{Black}
\Photon(40,30)(40,20){3}{2.5}
\ArrowLine(80,0)(60,0)
\ArrowLine(60,0)(20,0)
\ArrowLine(20,0)(0,0)
\Gluon(40,20)(20,0){3}{4.5}
\Gluon(40,20)(60,0){3}{4.5}
\Gluon(40,20)(40,0){3}{3.5}
\Vertex(40,20){3}
\LongArrow(48,30)(48,23)
\LongArrow(17,-5)(4,-5)
    \Text(38,20)[rb]{{\Black{$\tilde{O}_1^B$}}}
    \Text(52,22)[lb]{{\Black{$q$}}}
    \Text(5,-15)[lb]{{\Black{$q$}}}
  \end{picture}
+\ldots\right)$\\
$+Z_{GJ}\left(
 \begin{picture}(80,30) (0,0)
    \SetWidth{0.5}
    \SetColor{Black}
\Photon(40,30)(40,20){3}{2.5}
\ArrowLine(40,20)(0,0)
\ArrowLine(80,0)(40,20)
\Vertex(40,20){3}
\LongArrow(48,30)(48,23)
\LongArrow(18,2)(4,-5)
    \Text(38,20)[rb]{{\Black{$\partial_\nu J_5^{B\,\nu}$}}}
    \Text(52,22)[lb]{{\Black{$q$}}}
    \Text(17,-8)[lb]{{\Black{$q$}}}
  \end{picture}\,+
 \begin{picture}(80,30) (0,0)
    \SetWidth{0.5}
    \SetColor{Black}
\Photon(40,30)(40,20){3}{2.5}
\ArrowLine(40,20)(20,10)
\ArrowLine(20,10)(0,0)
\ArrowLine(80,0)(60,10)
\ArrowLine(60,10)(40,20)
\Gluon(60,10)(20,10){3}{4.5}
\Vertex(40,20){3}
\LongArrow(48,30)(48,23)
\LongArrow(18,2)(4,-5)
    \Text(38,20)[rb]{{\Black{$\partial_\nu J_5^{B\,\nu}$}}}
    \Text(52,22)[lb]{{\Black{$q$}}}
    \Text(17,-8)[lb]{{\Black{$q$}}}
  \end{picture}
\right)=\text{finite}
$\\[4ex]
\caption{Diagrams for the calculation of $Z_{GJ}$} \label{ZGJdias}
\end{center}\end{figure}

The $\eps$-tensors appearing in \re{O1t} and \re{dJ5Larin} are then drawn out of the R-operation performed in dimensional regularization.
In the correlators which have to be calculated there are always two $\eps$-tensors involved which can be contracted and expressed
through metric tensors:
\be \eps^{\mu_1\mu_2\mu_3\mu_4} \eps_{\nu_1\nu_2\nu_3\nu_4}=
-g^{[\mu_1}_{\,\,\,\nu_1} g^{\mu_2}_{\,\,\,\nu_2} g^{\mu_3}_{\,\,\,\nu_3} g^{\mu_4]}_{\,\,\,\nu_4}{},
\ee
where $[\ldots]$ means complete antisymmetrization.
These operators are renormalized like \cite{Larin:1993tq}
\bea
\,
\,[\partial_\mu J_5^\mu] &=& Z_5^s Z^s_{MS} \partial_\mu J_5^{B\,\mu}=Z_J^s \partial_\mu J_5^{B\,\mu},\label{renO1sdJ5_O1s}\\
\,[\tilde{O}_1] &=& Z_{G\tilde{G}} \tilde{O}_1^B+Z_{GJ} \partial_\mu J_5^{B\,\mu}{},
\label{renO1sdJ5_dJ5}
\eea
where $Z^s_{MS}$ is an $\overline{\text{MS}}$ renormalization constant, $Z_5^s$ a finite renormalization constant fixed by the
requirement that the one-loop character of the axial anomaly relation
\be
[\partial_\mu J_5^\mu]=\f{\alpha_s}{4\pi} \Nf \tr [\tilde{O}_1]+\text{CT} \label{O1sdJ5relation}
\ee
is valid in dimensional regularization.\footnote{In Pauli-Villars regularization for example this relation is automatically fulfilled.
In $d\neq 4$ dimensions, however, the operators $\partial_\mu J_5^\mu$ and $\tilde{O}_1$ become linearly independent.}
CT stands for contact terms of $\partial_\mu J_5^\mu$ with fermion fields. In the gluon sector these can be neglected.
$Z_{G\tilde{G}}$ is an $\overline{\text{MS}}$ renormalization constant again and $Z_{GJ}$ starts at $\mathcal{O}(\als)$.
In \cite{Larin:1993tq} $Z^s_{MS}$ and $Z_5^s$ are given up to $\mathcal{O}(\als^3)$ and $\mathcal{O}(\als^2)$ respectively.
Furthermore it is shown that $Z_{G\tilde{G}}=Z_a$ ($Z_a$ being the renormalization constant for $\als$). The constant $Z_{GJ}$ is only
given at one-loop level in the literature \cite{Larin:1993tq, Chetyrkin:1998mw} but for the Wilson coefficient $C_1$ at three-loop level
it is needed to two-loop accuracy. In section \ref{chapRGIsection} we will also need the corresponding three-loop anomalous dimension.
The simplest way to determine $Z_{GJ}$ is by constructing the matrix elements of $\tilde{O}_1$ and
$\partial_\mu J_5^{B\,\mu}$ with two external fermions (see Fig. \ref{ZGJdias}) using a projector
\be P(q):= q^{\mu_1}\gamma^{\mu_2}\gamma^{\mu_3}\gamma^{\mu_4} \eps_{\mu_1\mu_2\mu_3\mu_4} \ee on the external fermion line.
From this we get
\be \begin{split} 
Z_{GJ}&=\f{\als}{4\pi\eps} 12 \cf 
+ \f{\als^2}{(4\pi)^2\eps}\left\{\f{142\ca\cf}{3}-42\cf^2-\f{8}{3}\cf\Nf\tr\right\}\\
&+\f{\als^2}{(4\pi)^2\eps^2}\left\{16\cf\Nf\tr-44\ca\cf\right\}\\ &
 + \f{\als^3}{(4\pi)^3\eps^3} \left\{
           \f{484}{3} \ca^2 \cf
          - \f{352}{3} \Nf \ca \tr \cf
          + \f{64}{3} \Nf^2 \tr^2 \cf
          \right\}\\ &
       + \f{\als^3}{(4\pi)^3\eps^2} \left\{
           \f{550}{3} \ca \cf^2
          - \f{2378}{9} \ca^2 \cf
          - \f{32}{3} \Nf \tr \cf^2 \right. \\ & \left.
          + \f{1136}{9} \Nf \ca \tr \cf
          - \f{32}{9} \Nf^2 \tr^2 \cf
          \right\}\\ &
       + \f{\als^3}{(4\pi)^3\eps} \left\{
           178 \cf^3
          - \f{2947}{9} \ca \cf^2
          + \f{1607}{9} \ca^2 \cf
          - \f{1096}{9} \Nf \tr \cf^2 \right. \\ & \left.
          + \f{328}{9} \Nf \ca \tr \cf 
          - \f{208}{9} \Nf^2 \tr^2 \cf
          + 192 \zeta_{3} \Nf \tr \cf^2
          - 192 \zeta_{3} \Nf \ca \tr \cf
          \right\}{}.
\end{split} \label{ZGJ}
\ee
An interesting additional application of this result is to check the connection between the anomalous dimensions
of the operator set $\{\tilde{O}_1,\partial_\nu J_5^{\nu}\}$. In \cite{Larin:1993tq} the following relations have been motivated:
\bea \gamma_{G\tilde{G}} &=& -\f{\beta(\als)}{\als} \label{relgGGtbeta}{},\\
     \gamma_{GJ} &=& \lb \f{\als}{4\pi} \Nf \tr \rb^{-1} \gamma_J^s{}, \label{relgGJgJs}
\eea
with \be \gamma_{ij}=\left(\mu^2\f{d}{d\mu^2}Z_{ik}\right) \left( Z^{-1} \right)_{kj}, \quad Z=\left(
\begin{array}{cc}
 Z_{G\tilde{G}} & Z_{GJ} \\ 0 & Z_J^s
\end{array} \right).
 \label{gammadef}\ee 
The first relation \re{relgGGtbeta} has been explicitly checked to three-loop
level in \cite{Larin:1993tq} the second one \re{relgGJgJs} only to one-loop accuracy. Now we can check this equation with $\gamma_{GJ}$
at two-loop level and $\gamma_J^s$ at three-loop level and it turns out to hold there as well.
Using \re{ZGJ} and the renormalization constants $Z_J^s$ and $Z_a$ \cite{Larin:1993tq, Larin:1993tp} the following
anomalous dimension is derived:\footnote{$\gamma_{G\tilde{G}}$ and $\gamma_{J}^s$ can be found in \cite{Larin:1993tq, Larin:1993tp} 
at three-loop level. All renormalization constants and anomalous dimensions are available at
\texttt{\bf http://www-ttp.particle.uni-karlsruhe.de/Progdata/ttp13/ttp13-003/}}
\be \begin{split}
\gamma_{GJ}&=-12\cf  \left(\f{\als}{4\pi}\right) +\left(\f{\als}{4\pi}\right)^2 
\left\{-\f{284}{3}  \ca \cf +36  \cf^2 +\f{16}{3}  \cf \Nf \tr \right\}\\ &
+\left(\f{\als}{4\pi}\right)^3 \left\{
-\f{1607}{3}  \ca^2 \cf
+461  \ca \cf^2
+576  \ca \cf \Nf \tr \zeta_3
-\f{328}{3}  \ca \cf \Nf \tr \right. \\ & \left.
-126  \cf^3
-576  \cf^2 \Nf \tr \zeta_3
+428  \cf^2 \Nf \tr
+\f{208}{3}  \cf \Nf^2 \tr^2\right\}.
\end{split} \ee

Now we can write the correlator $X_t(q)$ as
\be \begin{split}
 &i\int\!\mathrm{d}^4x\,e^{iqx} T\{\,[\tilde{O}_1](x)[\tilde{O}_1](0)\}\\
=&i\int\!\mathrm{d}^4x\,e^{iqx} T\left\{Z_{G\tilde{G}}^2
\tilde{O}^B_1(x)\tilde{O}^B_1(0) + 2 Z_{G\tilde{G}} Z_{GJ}
\tilde{O}^B_1(x)\partial_\mu J_5^{B\,\mu}(0) + Z_{GJ}^2
\partial_\mu J_5^{B\,\mu}(x) \partial_\nu J_5^{B\,\nu}(0) \right\}{}.    
    \end{split} \label{O1sO1s} \ee

In \cite{Zoller:2012qv} it has been discovered that there are contact terms at two-loop level in the coefficient $C_1$
for the correlator of $O_1$. The coefficient $C_0$ also has contact terms for the correlator of two operators $O_1$ 
or two operators $T^{\mu\nu}$.
For the operator $\tilde{O}_1$ we can make an important restriction on possible contact terms due to the fact that
it can be exactly expressed as the divergence of the Chern-Simons current:
\be
\tilde{O}_1=\partial_\mu K^\mu \label{divChernSimons}
\ee
with
\be
K^\mu=\eps^{\mu\nu\rho\sigma} \left\{4 G^a_\nu \partial_\rho G^a_\sigma + \f{4}{3}\gs f^{abc} G^a_\nu G^b_\rho  G^c_\sigma\right\}.
\label{ChernSimons}
\ee
From this follows for \re{O1sO1s}
\be \begin{split}
 & i\int\!\mathrm{d}^4x\,e^{iqx} T\{[\tilde{O}_1](x)[\tilde{O}_1](0)\}\\
=& q_\mu q_\nu\,   i\int\!\mathrm{d}^4x\,e^{iqx} T\left\{Z_{G\tilde{G}}^2
K^{B\,\mu}(x)K^{B\,\nu}(0) + 2 Z_{G\tilde{G}} Z_{GJ}
K^{B\,\mu}(x)J_5^{B\,\nu}(0) + Z_{GJ}^2
J_5^{B\,\mu}(x) J_5^{B\,\nu}(0) \right\}\\
\rightarrow& q_\mu q_\nu \left\{q^2 C_0^{\mu\nu}(q^2)+ \f{1}{q^2}C_1^{\mu\nu}(q^2) +\ldots\right\}
\text{ for }q^2\rightarrow -\infty \text{ (OPE)}
    \end{split} \label{O1sO1swithK} \ee
with dimensionless coefficients $C_0^{\mu\nu}(q^2)$ and $C_1^{\mu\nu}(q^2)$. Because of the non-local factor $\f{1}{q^2}$
the coefficient $C_1^{\mu\nu}(q^2)$ cannot contain any contact terms. This makes the Wilson coefficent
$C_1(q^2)=\f{q_\mu q_\nu}{q^2}C_1^{\mu\nu}(q^2)$ for the correlator \re{O1sO1s} finite and unambiguous due to the absence of
contact terms.
\section{Calculation and results} \label{chapCalcandRes}
\begin{figure}[h!]
\begin{center}
$
\left.\langle 0|X_t(q)|0 \rangle\right|_{\text{pert}}=$\\[2ex]
$Z_{a}^2\left(
  \begin{picture}(80,30) (0,0)
    \SetWidth{0.5}
    \SetColor{Black}
\Photon(5,0)(20,0){3}{2.5}
\Photon(60,0)(75,0){3}{2.5}
\LongArrow(7,8)(15,8)
\LongArrow(65,8)(73,8)
\CCirc(40,0){20}{Black}{Blue}
    \Text(0,-17)[lb]{{\Black{$\tilde{O}_1^B$}}}
    \Text(62,-17)[lb]{{\Black{$\tilde{O}_1^B$}}}
    \Text(7,12)[lb]{{\Black{$q$}}}
    \Text(64,12)[lb]{{\Black{$q$}}}
  \end{picture}\right)
+2Z_{a} Z_{GJ}\left(
  \begin{picture}(101,30) (0,0)
    \SetWidth{0.5}
    \SetColor{Black}
\Photon(5,0)(20,0){3}{2.5}
\Photon(60,0)(75,0){3}{2.5}
\LongArrow(7,8)(15,8)
\LongArrow(65,8)(73,8)
\CCirc(40,0){20}{Black}{Blue}
    \Text(0,-17)[lb]{{\Black{$\tilde{O}_1^B$}}}
    \Text(62,-17)[lb]{{\Black{$\partial_\nu J_5^{B\,\nu}$}}}
    \Text(7,12)[lb]{{\Black{$q$}}}
    \Text(64,12)[lb]{{\Black{$q$}}}
  \end{picture}\right)
+Z_{GJ}^2\left(
  \begin{picture}(101,30) (-10,0)
    \SetWidth{0.5}
    \SetColor{Black}
\Photon(5,0)(20,0){3}{2.5}
\Photon(60,0)(75,0){3}{2.5}
\LongArrow(7,8)(15,8)
\LongArrow(65,8)(73,8)
\CCirc(40,0){20}{Black}{Blue}
    \Text(-10,-17)[lb]{{\Black{$\partial_\nu J_5^{B\,\nu}$}}}
    \Text(62,-17)[lb]{{\Black{$\partial_\nu J_5^{B\,\nu}$}}}
    \Text(7,12)[lb]{{\Black{$q$}}}
    \Text(64,12)[lb]{{\Black{$q$}}}
  \end{picture}\right)$\\[4ex]
$=Z_{G\tilde{G}}^2\left(
  \begin{picture}(90,30) (-1,0)
    \SetWidth{0.5}
    \SetColor{Black}
\Photon(5,0)(20,0){3}{2.5}
\Photon(60,0)(75,0){3}{2.5}
\GlueArc(40,0)(20,0,180){3}{9.5}
\GlueArc(40,0)(20,180,360){3}{9.5}
    \Vertex(20,0){3}
    \Vertex(60,0){3}
    \Text(0,-17)[lb]{{\Black{$\tilde{O}_1^B$}}}
    \Text(62,-17)[lb]{{\Black{$\tilde{O}_1^B$}}}
  \end{picture}
+
\begin{picture}(90,30) (-1,0)
    \SetWidth{0.5}
    \SetColor{Black}
\Photon(5,0)(20,0){3}{2.5}
\Photon(60,0)(75,0){3}{2.5}
\GlueArc(40,0)(20,0,90){3}{4.5}
\GlueArc(40,0)(20,90,180){3}{4.5}
\GlueArc(40,0)(20,180,270){3}{4.5}
\GlueArc(40,0)(20,270,360){3}{4.5}
\Gluon(40,20)(40,-20){3}{6.5}
    \Vertex(40,20){2}
    \Vertex(40,-20){2}
    \Vertex(20,0){3}
    \Vertex(60,0){3}
    \Text(0,-17)[lb]{{\Black{$\tilde{O}_1^B$}}}
    \Text(62,-17)[lb]{{\Black{$\tilde{O}_1^B$}}}
\end{picture}
+
  \begin{picture}(90,30) (-1,0)
    \SetWidth{0.5}
    \SetColor{Black}
\Photon(5,0)(20,0){3}{2.5}
\Photon(60,0)(75,0){3}{2.5}
\GlueArc(40,0)(20,0,90){3}{4.5}
\GlueArc(40,0)(20,90,180){3}{4.5}
\GlueArc(40,0)(20,180,270){3}{4.5}
\GlueArc(40,0)(20,270,360){3}{4.5}
\Gluon(40,20)(40,10){3}{1}
\Gluon(40,-10)(40,-20){3}{1}
    \Vertex(40,20){2}
    \Vertex(40,-20){2}
    \Vertex(40,10){2}
    \Vertex(40,-10){2}
\ArrowArc(40,0)(10,90,270)
\ArrowArc(40,0)(10,270,90)
    \Vertex(20,0){3}
    \Vertex(60,0){3}
    \Text(0,-17)[lb]{{\Black{$\tilde{O}_1^B$}}}
    \Text(62,-17)[lb]{{\Black{$\tilde{O}_1^B$}}}
  \end{picture}
+
  \begin{picture}(90,30) (-1,0)
    \SetWidth{0.5}
    \SetColor{Black}
\Photon(5,0)(20,0){3}{2.5}
\Photon(60,0)(75,0){3}{2.5}
\GlueArc(40,0)(20,0,90){3}{4.5}
\GlueArc(40,0)(20,90,180){3}{4.5}
\GlueArc(40,0)(20,180,270){3}{4.5}
\GlueArc(40,0)(20,270,360){3}{4.5}
\Gluon(40,20)(40,10){3}{1}
\Gluon(40,-10)(40,-20){3}{1}
    \Vertex(40,20){2}
    \Vertex(40,-20){2}
    \Vertex(40,10){2}
    \Vertex(40,-10){2}
\GlueArc(40,0)(10,90,270){3}{4.5}
\GlueArc(40,0)(10,270,90){3}{4.5}
    \Vertex(20,0){3}
    \Vertex(60,0){3}
    \Text(0,-17)[lb]{{\Black{$\tilde{O}_1^B$}}}
    \Text(62,-17)[lb]{{\Black{$\tilde{O}_1^B$}}}
  \end{picture}+\ldots\right)$\\[4ex]
$+2 Z_{G\tilde{G}} Z_{GJ}\left(
\begin{picture}(101,30) (-10,0)
    \SetWidth{0.5}
    \SetColor{Black}
\Photon(5,0)(20,0){3}{2.5}
\Photon(60,0)(75,0){3}{2.5}
\ArrowArc(40,0)(20,0,90)
\GlueArc(40,0)(20,90,180){3}{4.5}
\GlueArc(40,0)(20,180,270){3}{4.5}
\ArrowArc(40,0)(20,270,360)
\ArrowLine(40,20)(40,-20)
    \Vertex(40,20){2}
    \Vertex(40,-20){2}
    \Vertex(20,0){3}
    \Vertex(60,0){3}
    \Text(0,-17)[lb]{{\Black{$\tilde{O}_1^B$}}}
    \Text(62,-17)[lb]{{\Black{$\partial_\nu J_5^{B\,\nu}$}}}
\end{picture}\right)
+Z_{GJ}^2\left(
  \begin{picture}(101,30) (-10,0)
    \SetWidth{0.5}
    \SetColor{Black}
\Photon(5,0)(20,0){3}{2.5}
\Photon(60,0)(75,0){3}{2.5}
\ArrowArc(40,0)(20,0,180)
\ArrowArc(40,0)(20,180,360)
    \Vertex(20,0){3}
    \Vertex(60,0){3}
    \Text(-10,-17)[lb]{{\Black{$\partial_\nu J_5^{B\,\nu}$}}}
    \Text(62,-17)[lb]{{\Black{$\partial_\nu J_5^{B\,\nu}$}}}
  \end{picture}\right)$\\[7ex]
\caption{Diagrams for the calculation of the coefficient $C_0(Q^2)$} \label{O0_dias}
\end{center}
\end{figure}
\subsection{Details of the calculation} \label{chapSetup}
The leading coefficient $C_0$ is just the perturbative VEV of the correlator eq.~\re{O1sO1s} 
\be
(Q^2)^2 C_0(q)=\left.\langle 0|i\int\!\mathrm{d}^4x\,e^{iqx} T\{[\tilde{O}_1](x)[\tilde{O}_1](0)\}|0\rangle\right|_{\text{pert}}
\ee
which has been computed up to order $\als^2$ (three loops).  In Figure
\re{O0_dias} some sample Feynman diagrams contributing to
this calculation are shown. The operators $\tilde{O}_1^B$ and $\partial_\mu J_5^{B\,\mu}$ play the roles of
external currents. The Feynman diagrams have been produced with
the program QGRAF \cite{QGRAF}. As all diagrams in this problem are
propagator-like the relevant integrals can be computed with the FORM
package MINCER \cite{Vermaseren:2000nd,Larin:1991fz,MINCER}. For the colour part of the diagrams the FORM package COLOR
\cite{vanRitbergen:1998pn} has been used.

In order to compute the coefficient $C_1(Q^2)$
the method of projectors \cite{Gorishnii:1983su,Gorishnii:1986gn} has been applied, which allows to
express coefficient functions for any OPE of two operators in terms of
massless propagator type diagrams only. 
The method is based the fact that in dimensional regularization every massless tadpole-like Feynman integral is set to zero.\\
We apply a projector to both sides of \re{OPE_B} which sets every operator on the rhs to zero except for $O_1^B$:
\be 
{\bf P}\{X_t(q)\}=\sum \limits_i (Q^2)^\f{4-\text{dim}(O_{i})}{2}\, C_{i}^{B,(r)}(Q^2)\, {\bf P}\{O_{i}^B\}{},
 \label{Xtproj}
\ee
with ${\bf P}\{O_{1}^B\}=1$ and ${\bf P}\{O_{i\neq 1}^B\}=0$.
This is done in the same way as described in \cite{Zoller:2012qv} leading to
\be
C_{1,B}(Q^2)=Z_{G\tilde{G}}^2 C_{1,B}^{(\tilde{O}_1^B, \tilde{O}_1^B)}(Q^2)
+2 Z_{G\tilde{G}} Z_{GJ} C_{1,B}^{(\tilde{O}_1^B, \partial_\nu J_5^{B\,\nu})}(Q^2)
+Z_{GJ}^2 C_{1,B}^{(\partial_\nu J_5^{B\,\nu}, \partial_\nu J_5^{B\,\nu})}(Q^2),
\ee
with
\be
C_{1,B}^{(O^B_\alpha O^B_\beta)}(Q^2) =\f{\delta^{ab}}{\Ng}\f{g^{\mu_1 \mu_2}}{(D-1)}
\f{1}{D}\f{\p}{\p k_1} \cd \f{\p}{\p k_2} \left. \left[
  \begin{picture}(165,50) (0,0)
    \SetWidth{0.5}
    \SetColor{Black}
    \Gluon(10,0)(50,0){5.5}{4.5}
    \Gluon(110,0)(150,0){5.5}{4.5}
    \LongArrow(35,15)(25,15)
    \LongArrow(125,15)(135,15)
\Photon(80,0)(120,40){3}{4}
\Photon(80,0)(40,40){3}{4}
    \CCirc(80,0){30}{Black}{Blue}
    \SetColor{Red}
\LongArrowArc(80,73)(40,220,320)
\Vertex(110,0){4}
\Vertex(50,0){4}
    \SetColor{Black}
    \Text(25,17)[lb]{\Large{\Black{$k_1$}}}
    \Text(125,17)[lb]{\Large{\Black{$k_2$}}}
    \Text(36,-17)[lb]{\Large{\Red{$g_B$}}}
    \Text(110,-17)[lb]{\Large{\Red{$g_B$}}}
    \Text(80,40)[c]{\Large{\Red{$q$}}}
    \Text(150,-20)[lb]{\Large{\Black{$\mu_2$}}}
    \Text(5,-20)[lb]{\Large{\Black{$\mu_1$}}}
    \Text(5,10)[lb]{\Large{\Black{$a$}}}
    \Text(150,10)[lb]{\Large{\Black{$b$}}}
    \Text(20,35)[lb]{\Large{\Black{$O^B_\alpha$}}}
    \Text(125,35)[lb]{\Large{\Black{$O^B_\beta$}}}
  \end{picture}
\right] \right|_{k_i=0}
{},
\ee
where the blue circle represents the the sum of all (bare) Feynman diagrams
which become 1PI after formal gluing of the two external lines representing the operators on the lhs of the OPE. 
\begin{figure}[h!] 
\begin{center}
\vskip 2ex
$  \begin{picture}(80,30) (0,0)
    \SetWidth{0.5}
    \SetColor{Black}
\Photon(5,-15)(20,0){3}{2}
\Photon(60,0)(75,-15){3}{2}
\Gluon(20,0)(60,0){3}{6.5}
\Gluon(20,0)(5,25){3}{4.5}
\Gluon(60,0)(75,25){3}{4.5}
\LongArrowArcn(40,-20)(15,160,25)
    \Vertex(20,0){3}
    \Vertex(60,0){3}
    \Text(-3,-7)[lb]{{\Black{$\tilde{O}_1^B$}}}
    \Text(67,-7)[lb]{{\Black{$\tilde{O}_1^B$}}}
    \Text(35,-20)[lb]{{\Black{$q$}}}
  \end{picture}   \; + \;
\begin{picture}(80,30) (0,0)
    \SetWidth{0.5}
    \SetColor{Black}
\Photon(5,0)(20,0){3}{2}
\Photon(60,0)(75,0){3}{2}
\GluonArc(40,0)(20,180,270){3}{4.5}
\GluonArc(40,0)(20,90,180){3}{4.5}
\GluonArc(40,0)(20,0,90){3}{4.5}
\GluonArc(40,0)(20,270,360){3}{4.5}
\Gluon(40,20)(40,40){3}{3.5}
\Gluon(40,-20)(40,-40){3}{3.5}
    \Vertex(20,0){3}
    \Vertex(60,0){3}
  \end{picture}
 \; + \;
\begin{picture}(80,30) (0,0)
    \SetWidth{0.5}
    \SetColor{Black}
\Photon(5,0)(20,0){3}{2}
\Photon(60,0)(75,0){3}{2}
\GluonArc(40,0)(20,180,270){3}{4.5}
\GluonArc(40,0)(20,90,180){3}{4.5}
\GluonArc(40,0)(20,0,90){3}{4.5}
\GluonArc(40,0)(20,270,360){3}{4.5}
\Gluon(40,20)(40,40){3}{3.5}
\Gluon(40,-20)(40,-40){3}{3.5}
\Gluon(20,0)(60,0){3}{6.5}
    \Vertex(20,0){3}
    \Vertex(60,0){3}
  \end{picture}
 \; + \;
\begin{picture}(80,30) (0,0)
    \SetWidth{0.5}
    \SetColor{Black}
\Photon(5,-5)(20,0){3}{2}
\Photon(60,0)(75,-5){3}{2}
\GluonArc(40,0)(20,180,360){3}{8.5}
\GluonArc(40,0)(20,110,180){3}{3.5}
\GluonArc(40,0)(20,0,70){3}{3.5}
\ArrowArc(40,20)(7,0,180)
\ArrowArc(40,20)(7,180,360)
\Gluon(20,0)(60,0){3}{6.5}
\Gluon(20,0)(5,20){3}{3.5}
\Gluon(60,0)(75,20){3}{3.5}
    \Vertex(20,0){3}
    \Vertex(60,0){3}
  \end{picture}
\,+\,\ldots$\\[7ex]
\caption{Diagrams for the calculation of $C_{1,B}^{(\tilde{O}_1^B, \tilde{O}_1^B)}$.} \label{O1sO1s_dias}
\end{center}
\end{figure}
\begin{figure}[h!] 
\begin{center}
\vskip 2ex
$\begin{picture}(80,30) (0,0)
    \SetWidth{0.5}
    \SetColor{Black}
\Photon(5,-15)(20,0){3}{2}
\Photon(60,0)(75,-15){3}{2}
\Gluon(40,0)(60,0){3}{3.5}
\ArrowArc(30,0)(10,180,0)
\ArrowArc(30,0)(10,0,90)
\ArrowArc(30,0)(10,90,180)
\Gluon(30,10)(30,30){3}{3.5}
\Gluon(60,0)(60,30){3}{4.5}
    \Vertex(20,0){3}
    \Vertex(60,0){3}
\LongArrowArcn(40,-28)(15,160,25)
    \Text(-15,-7)[lb]{{\Black{$\partial_\nu J_5^{B\,\nu}$}}}
    \Text(67,-7)[lb]{{\Black{$\tilde{O}_1^B$}}}
    \Text(35,-25)[lb]{{\Black{$q$}}}
  \end{picture}\quad+\,
\begin{picture}(80,30) (0,0)
    \SetWidth{0.5}
    \SetColor{Black}
\Photon(5,-5)(20,0){3}{2}
\Photon(60,0)(75,-5){3}{2}
\Gluon(40,0)(60,0){3}{3.5}
\ArrowArc(30,0)(10,180,0)
\ArrowArc(30,0)(10,0,90)
\ArrowArc(30,0)(10,90,180)
\GluonArc(45,0)(15,220,360){3}{6.5}
\Gluon(30,10)(30,30){3}{3.5}
\Gluon(60,0)(60,30){3}{4.5}
    \Vertex(20,0){3}
    \Vertex(60,0){3}
  \end{picture}\;+\,\ldots$
\\[6ex]
\caption{Diagrams for the calculation of $C_{1,B}^{(\tilde{O}_1^B, \partial_\nu J_5^{B\,\nu})}$.} \label{O1sdJ5_dias}
\end{center}
\end{figure}
\begin{figure}[h!] 
\begin{center}
\vskip 2ex
$
\begin{picture}(80,30) (0,0)
    \SetWidth{0.5}
    \SetColor{Black}
\Photon(5,0)(20,0){3}{2}
\Photon(60,0)(75,0){3}{2}
\ArrowArc(40,0)(20,180,270)
\ArrowArc(40,0)(20,90,180)
\ArrowArc(40,0)(20,0,90)
\ArrowArc(40,0)(20,270,360)
\Gluon(40,20)(40,40){3}{3.5}
\Gluon(40,-20)(40,-40){3}{3.5}
    \Vertex(20,0){3}
    \Vertex(60,0){3}
\LongArrow(0,10)(10,10)
\LongArrow(70,10)(80,10)
    \Text(-13,-16)[lb]{{\Black{$\partial_\nu J_5^{B\,\nu}$}}}
    \Text(67,-16)[lb]{{\Black{$\partial_\nu J_5^{B\,\nu}$}}}
    \Text(2,20)[lb]{{\Black{$q$}}}
    \Text(70,20)[lb]{{\Black{$q$}}}
  \end{picture}
\qquad+\,
\begin{picture}(80,30) (0,0)
    \SetWidth{0.5}
    \SetColor{Black}
\Photon(5,0)(20,0){3}{2}
\Photon(60,0)(75,0){3}{2}
\ArrowArc(40,0)(20,120,180)
\ArrowArc(40,0)(20,60,120)
\ArrowArc(40,0)(20,0,60)
\ArrowArc(40,0)(20,180,360)
\Gluon(30,17)(30,40){3}{3.5}
\Gluon(50,17)(50,40){3}{3.5}
    \Vertex(20,0){3}
    \Vertex(60,0){3}
  \end{picture}
\;+\,\ldots$\\[7ex]
\caption{Diagrams for the calculation of $C_{1,B}^{(\partial_\nu J_5^{B\,\nu}, \partial_\nu J_5^{B\,\nu})}$.} \label{dJ5dJ5_dias} 
\end{center}
\end{figure}

Table \re{numdias} shows the number of diagrams generated for the different contributions to $C_0$ and $C_1$.
Sample diagrams for the calculation of the bare coefficients $C_{1,B}^{(\tilde{O}_1^B, \tilde{O}_1^B)}$,
$C_{1,B}^{(\tilde{O}_1^B, \partial_\nu J_5^{B\,\nu})}$ and
$C_{1,B}^{(\partial_\nu J_5^{B\,\nu}, \partial_\nu J_5^{B\,\nu})}$ are shown in Figures \re{O1sO1s_dias}, \re{O1sdJ5_dias} and \re{dJ5dJ5_dias} respectively.

\begin{table}[h!]
\begin{center}
\begin{tabular}{|l|llll|}
\hline
Correlator & 0 loop & 1 loop & 2 loop & 3 loop \\
\hline
$\langle 0|\tilde{O}_1^B(x)\tilde{O}_1^B(0)|0\rangle_{\text{pert}}$ & 0 & 1 & 12 & 215 \\
$\langle 0|\tilde{O}_1^B(x)\partial_\nu J_5^{B\,\nu}(0)|0\rangle_{\text{pert}}$ & 0 & 0 & 1 & \\
$\langle 0|\partial_\nu J_5^{B\,\nu}(x)\partial_\nu J_5^{B\,\nu}(0)|0\rangle_{\text{pert}}$ & 0 & 1 & & \\
${\bf P_1}(\tilde{O}_1^B(x)\tilde{O}_1^B(0))$ & 2 & 75 & 2567 & 94964 \\
${\bf P_1}(\tilde{O}_1^B(x)\partial_\nu J_5^{B\,\nu}(0))$ & 0 & 8 & 345 & \\
${\bf P_1}(\partial_\nu J_5^{B\,\nu}(x)\partial_\nu J_5^{B\,\nu}(0))$ & 0 & 8 & & \\
\hline
\end{tabular}
\end{center}
\caption{Number of diagrams needed for $C_0$ and $C_1$}
\label{numdias}
\end{table}
All results are given in the $\overline{\text{MS}}$ scheme with
$\as=\f{\als}{\pi}$, $\als=\f{\gs^2}{4\pi}$ and the
abbreviation $l_{\sss{\mu q}}=\ln\lb\f{\mu^2}{Q^2}\rb$ where $\mu$ is the
$\overline{\text{MS}}$ renormalization scale.
They can be retrieved from\\
\texttt{\bf http://www-ttp.particle.uni-karlsruhe.de/Progdata/ttp13/ttp13-003/}

The gauge group factors are defined in the usual way: $\cf$ and $\ca$ are the quadratic Casimir
operators of the quark and the adjoint representation of the corresponding Lie algebra,
$\dR$ is the dimension of the quark representation, $\Ng$ is the number of gluons (dimension of the adjoint representation),
$\tr$ is defined so that \mbox{$\tr \delta^{ab}=\textbf{Tr}\lb T^a T^b\rb$}  
is the trace of two group generators of the quark representation.\footnote{For an SU$(N)$ gauge group these are $\dR=N$,
$\ca=2\tr N$ and $\cf=\tr\lb N-\f{1}{N}\rb$.}
For  QCD (colour gauge group SU$(3)$) we have $\cf =4/3\,,\, \ca=3\,,\,\tr=1/2$ and $\dR = 3$.
By $\Nf$ we denote the number of active quark flavours.

\subsection{Results} \label{chapResults}
As we have seen from \re{O1sO1s} contact terms in $C_0$ are possible and it turns out that they appear starting from one loop.
Because of these contact terms an unambiguous result for $C_0$ can only be given up to local
(that is q-independent) contributions. To avoid the ambiguity the $Q^2$-derivative is presented:
\be \begin{split}
Q^2\f{d}{dQ^2}\,C_0=&
       \f{\Ng}{\pi^2}\left[       - 1
       + \as   \left(
          - \f{97}{12} \ca
          + \f{7}{3} \Nf \tr
          \right)
       + \as l_{\sss{\mu q}}   \left(
          - \f{11}{6} \ca
          + \f{2}{3} \Nf \tr
          \right) \right.  \\ &\left. 
       + \as^2   \left(
          - \f{51959}{864} \ca^2
          + \f{107}{12} \Nf \tr \cf
          + \f{3793}{108} \Nf \ca \tr
          - \f{251}{54} \Nf^2 \tr^2 \right.\right.  \\ &\left. \left. 
          + \f{55}{8} \zeta_{3} \ca^2
          - 3 \zeta_{3} \Nf \tr \cf
          + \f{1}{2} \zeta_{3} \Nf \ca \tr
          \right) \right.  \\ &\left. 
       + \as^2 l_{\sss{\mu q}}   \left(
          - \f{1135}{48} \ca^2
          + 2 \Nf \tr \cf
          + \f{46}{3} \Nf \ca \tr
          - \f{7}{3} \Nf^2 \tr^2
          \right) \right.  \\ &\left. 
       + \as^2 l_{\sss{\mu q}}^2   \left(
          - \f{121}{48} \ca^2
          + \f{11}{6} \Nf \ca \tr
          - \f{1}{3} \Nf^2 \tr^2
          \right)
\right].
\end{split} \label{adlerC0T}
\ee
This result has been derived before \cite{Chetyrkin:1998mw} which serves as a nice check for the setup.
As discussed above the coefficient $C_1$ is unambiguous and is therefore given in full:
\beq
\begin{split}
C_1= & 64 \left\{
       1
       + \as   \left(
           \f{157}{36} \ca
          - \f{5}{9} \Nf \tr
          + \f{11}{12} l_{\sss{\mu q}} \ca
          - \f{1}{3} l_{\sss{\mu q}} \Nf \tr
          \right) \right. \\ &\left.
       + \as^2  \left( 
           \f{25945}{1296} \ca^2
          - \f{11}{2} \Nf \tr \cf
          - \f{4355}{648} \Nf \ca \tr
          + \f{25}{81} \Nf^2 \tr^2     
          + \f{1727}{216} l_{\sss{\mu q}} \ca^2     \right.\right. \\ &\left.\left.
          - \f{3}{2} l_{\sss{\mu q}} \Nf \tr \cf
          - \f{106}{27} l_{\sss{\mu q}} \Nf \ca \tr
          + \f{10}{27} l_{\sss{\mu q}} \Nf^2 \tr^2
          + \f{121}{144} l_{\sss{\mu q}}^2 \ca^2
          - \f{11}{18} l_{\sss{\mu q}}^2 \Nf \ca \tr \right.\right. \\ &\left.\left.
          + \f{1}{9} l_{\sss{\mu q}}^2 \Nf^2 \tr^2
          - \f{33}{8} \zeta_{3} \ca^2
          + 3 \zeta_{3} \Nf \tr \cf
          - \f{3}{2} \zeta_{3} \Nf \ca \tr
          \right) \right. \\ &\left.
       + \as^3   \left(
           \f{19360399}{186624} \ca^3
          + \f{461}{144} \Nf \tr \cf^2
          - \f{614501}{10368} \Nf \ca \tr \cf \right.\right. \\ &\left.\left.
          - \f{1857805}{31104} \Nf \ca^2 \tr  
          + \f{28981}{2592} \Nf^2 \tr^2 \cf 
          + \f{126415}{15552} \Nf^2 \ca \tr^2
          - \f{125}{729} \Nf^3 \tr^3 \right.\right. \\ &\left.\left.
          + \f{594247}{10368} l_{\sss{\mu q}} \ca^3  
          + \f{35}{32} l_{\sss{\mu q}} \Nf \tr \cf^2
          - \f{1623}{64} l_{\sss{\mu q}} \Nf \ca \tr \cf 
          - \f{68935}{1728} l_{\sss{\mu q}} \Nf \ca^2 \tr  \right.\right. \\ &\left.\left.
          + \f{105}{16} l_{\sss{\mu q}} \Nf^2 \tr^2 \cf
          + \f{6661}{864} l_{\sss{\mu q}} \Nf^2 \ca \tr^2
          - \f{25}{81} l_{\sss{\mu q}} \Nf^3 \tr^3 
          + \f{9779}{864} l_{\sss{\mu q}}^2 \ca^3  \right.\right. \\ &\left.\left.
          - \f{275}{96} l_{\sss{\mu q}}^2 \Nf \ca \tr \cf
          - \f{2795}{288} l_{\sss{\mu q}}^2 \Nf \ca^2 \tr 
          + \f{25}{24} l_{\sss{\mu q}}^2 \Nf^2 \tr^2 \cf
          + \f{61}{24} l_{\sss{\mu q}}^2 \Nf^2 \ca \tr^2 \right.\right. \\ &\left.\left.
          - \f{5}{27} l_{\sss{\mu q}}^2 \Nf^3 \tr^3
          + \f{1331}{1728} l_{\sss{\mu q}}^3 \ca^3
          - \f{121}{144} l_{\sss{\mu q}}^3 \Nf \ca^2 \tr 
          + \f{11}{36} l_{\sss{\mu q}}^3 \Nf^2 \ca \tr^2  \right.\right. \\ &\left.\left.
          - \f{1}{27} l_{\sss{\mu q}}^3 \Nf^3 \tr^3 
          + \f{55}{8} \zeta_{5} \ca^3
          - 15 \zeta_{5} \Nf \tr \cf^2
          + \f{15}{2} \zeta_{5} \Nf \ca \tr \cf
          + 5 \zeta_{5} \Nf \ca^2 \tr     \right.\right. \\ &\left.\left.
          - \f{6893}{144} \zeta_{3} \ca^3 
          + \f{145}{12} \zeta_{3} \Nf \tr \cf^2
          + \f{1291}{48} \zeta_{3} \Nf \ca \tr \cf
          - \f{349}{144} \zeta_{3} \Nf \ca^2 \tr   \right.\right. \\ &\left.\left.
          - \f{13}{2} \zeta_{3} \Nf^2 \tr^2 \cf
          + \f{121}{36} \zeta_{3} \Nf^2 \ca \tr^2
          - \f{363}{32} \zeta_{3} l_{\sss{\mu q}} \ca^3
          + \f{33}{4} \zeta_{3} l_{\sss{\mu q}} \Nf \ca \tr \cf   \right.\right. \\ &\left.\left.
          - 3 \zeta_{3} l_{\sss{\mu q}} \Nf^2 \tr^2 \cf
          + \f{3}{2} \zeta_{3} l_{\sss{\mu q}} \Nf^2 \ca \tr^2
        \right)
\right\}.
\label{C1:O1s}
\end{split}
\eeq
The cancellation of all divergences is a strong check for this result. Another important check is the independence of the gauge
parameter $\xi$ as all calculations have been done for an arbitrary $R_\xi$ gauge.
The leading term of \re{C1:O1s} is in agreement with \cite{Novikov:1979ux} and the part $\propto \as l_{\sss{\mu q}}$
has been derived in \cite{Zhang:2003mr} if we set the colour factors to their QCD values.\footnote{In 
\cite{Zhang:2003mr}, however, the leading term differs from this result
and the one derived in \cite{Novikov:1979ux} by a minus sign and the non-logarithmic term of $\mathcal{O}(\as)$ is also missing there.}
For QCD colour factors we get
\beq
\begin{split}
C_1= & 64\left\{
       1
       + a_s   \left(
           \f{157}{12}
          - \f{5}{18} \Nf
          + \f{11}{4} l_{\sss{\mu q}}
          - \f{1}{6} l_{\sss{\mu q}} \Nf
          \right) \right. \\ &\left.
       + a_s^2   \left(
           \f{25945}{144}
          - \f{5939}{432} \Nf
          + \f{25}{324} \Nf^2
          + \f{1727}{24} l_{\sss{\mu q}}  \right.\right. \\ &\left.\left.
          - \f{62}{9} l_{\sss{\mu q}} \Nf
          + \f{5}{54} l_{\sss{\mu q}} \Nf^2
          + \f{121}{16} l_{\sss{\mu q}}^2  \right.\right. \\ &\left.\left.
          - \f{11}{12} l_{\sss{\mu q}}^2 \Nf
          + \f{1}{36} l_{\sss{\mu q}}^2 \Nf^2
          - \f{297}{8} \zeta_{3}
          - \f{1}{4} \zeta_{3} \Nf
          \right) \right. \\ &\left.
       + a_s^3   \left(
           \f{19360399}{6912}
          - \f{7972411}{20736} \Nf
          + \f{611093}{62208} \Nf^2
          - \f{125}{5832} \Nf^3  \right.\right. \\ &\left.\left.
          + \f{594247}{384} l_{\sss{\mu q}}
          - \f{264113}{1152} l_{\sss{\mu q}} \Nf
          + \f{9181}{1152} l_{\sss{\mu q}} \Nf^2
          - \f{25}{648} l_{\sss{\mu q}} \Nf^3  \right.\right. \\ &\left.\left.
          + \f{9779}{32} l_{\sss{\mu q}}^2
          - \f{9485}{192} l_{\sss{\mu q}}^2 \Nf
          + \f{649}{288} l_{\sss{\mu q}}^2 \Nf^2
          - \f{5}{216} l_{\sss{\mu q}}^2 \Nf^3  \right.\right. \\ &\left.\left.
          + \f{1331}{64} l_{\sss{\mu q}}^3
          - \f{121}{32} l_{\sss{\mu q}}^3 \Nf
          + \f{11}{48} l_{\sss{\mu q}}^3 \Nf^2
          - \f{1}{216} l_{\sss{\mu q}}^3 \Nf^3  \right.\right. \\ &\left.\left.
          + \f{1485}{8} \zeta_{5}
          + \f{145}{6} \zeta_{5} \Nf
          - \f{20679}{16} \zeta_{3}
          + \f{46333}{864} \zeta_{3} \Nf  \right.\right. \\ &\left.\left.
          + \f{17}{48} \zeta_{3} \Nf^2
          - \f{9801}{32} \zeta_{3} l_{\sss{\mu q}}
          + \f{33}{2} \zeta_{3} l_{\sss{\mu q}} \Nf
          + \f{1}{8} \zeta_{3} l_{\sss{\mu q}} \Nf^2
          \right)
\right\}.
\label{C1:O1scf}
\end{split}
\eeq
A nice consistency check for these results is to perform an OPE of the correlator
\be
i\int\!\mathrm{d}^4x\,e^{iqx} T\{[\partial_\mu J_5^\mu](x)[\partial_\mu J_5^\mu](0)\}
=(Q^2)^2 C_0^{JJ}+C_1^{JJ} [O_1] +\ldots
\label{dJ5corr}
\ee
and then see that \re{O1sdJ5relation} is fulfilled (except for possible contact terms):
\bea
Q^2\f{d}{dQ^2}\,C_0^{JJ}=&\left(\f{\alpha_s}{4\pi} \Nf \tr \right)^2 Q^2\f{d}{dQ^2}\,C_0{}, \label{checkO1sdJ5relationC0}\\
C_1^{JJ}=&\left(\f{\alpha_s}{4\pi} \Nf \tr \right)^2 C_1{}. \label{checkO1sdJ5relationC1}\eea
Indeed we find
\be \begin{split}
Q^2\f{d}{dQ^2}\,C_0^{JJ}=& 
       \f{\Ng}{\pi^2} \left[ - \f{a_s^2 }{16} \Nf^2 \tr^2   \right]       
\end{split} \label{adlerC0JJ}
\ee
and
\be \begin{split}
C_1^{JJ}=& 4 a_s^2  \Nf^2 \tr^2\left\{
        1
       + a_s   \left(
           \f{157}{36}  \ca 
          - \f{5}{9} \Nf \tr
          + \f{11}{12} l_{\sss{\mu q}} \ca 
          - \f{1}{3} l_{\sss{\mu q}} \Nf \tr
          \right)
\right\}
\end{split} \label{C1JJ}
\ee
satisfying \re{checkO1sdJ5relationC0} and \re{checkO1sdJ5relationC1} 
up to the calculated accuracy of $\mathcal{O}(a_s^2)$ and $\mathcal{O}(a_s^3)$ respectively.

\subsection{RGI operators and Wilson coefficients} \label{chapRGIsection}
Note that the coefficients \re{adlerC0T} and \re{C1:O1s} are not Renormalization Group invariant (RGI). In this section
we take RGI versions of all operators and construct RGI Wilson coefficients.
For an operator that is renormalized multiplicatively like $\partial_\mu J_5^\mu$ in \re{renO1sdJ5_dJ5}
constructing a finite and RGI operator is straightforward (see e.g. \cite{Bos:1992nd}). Because of 
\be \mu^2\f{d}{d\mu^2}[\partial_\mu J_5^\mu]=\gamma_J^s(a_s(\mu)) [\partial_\mu J_5^\mu] \ee
we can define
\be [\partial_\mu J_5^\mu]^{\text{RGI}}:=\underbrace{\exp\left\{-\int\limits^{a_s(\mu)} \f{\gamma_J^s(a)}{a\,\beta(a)}da\right\}}_{=: E_2(a_s)}
[\partial_\mu J_5^\mu] \label{dJ5RGI}\ee
which fulfills $\mu^2\f{d}{d\mu^2}[\partial_\mu J_5^\mu]^{\text{RGI}}=0$. A remarkable
feature of the operator \re{dJ5RGI} is its renormalization scheme independence \cite{Chetyrkin:1993hk}.
If we start with a different renormalized operator \be [\partial_\mu J_5^\mu]':=Z(a_s)[\partial_\mu J_5^\mu] \ee we get 
\be {\gamma_J^s}'(a_s)= \gamma_J^s(a_s)+\mu^2\f{d}{d\mu^2}\ln(Z(a_s)) \ee which leads to 
\be E_2'(a_s)=\f{E_2(a_s)}{Z(a_s)} \ee and therefore to the same RGI operator
\be [\partial_\mu J_5^\mu]^{\text{RGI}}=E_2'(a_s)[\partial_\mu J_5^\mu]'=E_2(a_s)[\partial_\mu J_5^\mu]{}.\ee

If we apply the same procedure to the non-diagonal operator $\tilde{O}_1$ we get
an  RG {\em variant} operator
\be [\tilde{O}_1]^{\text{RGV}}:=
\underbrace{\exp\left\{-\int\limits^{a_s(\mu)} \f{\gamma_{G\tilde{G}}(a)}{a\,\beta(a)}da\right\}}_{=: E_1(a_s)}
[\tilde{O}_1]{}, \ee
where $E_1(a_s)=a_s$ because of \re{relgGGtbeta}. Taking the derivative wrt the renormalization scale we find
\be \mu^2\f{d}{d\mu^2} [\tilde{O}_1]^{\text{RGV}}=E_1(a_s) \gamma_{GJ}(a_s) [\partial_\mu J_5^\mu]=
\f{E_1(a_s)}{E_2(a_s)} \gamma_{GJ}(a_s)[\partial_\mu J_5^\mu]^{\text{RGI}} \ee
which leads to the definition of the RGI operator
\be \begin{split}
     [\tilde{O}_1]^{\text{RGI}}:=&[\tilde{O}_1]^{\text{RGV}}-\underbrace{\int\limits^{a_s(\mu)}\f{E_1(a)}{E_2(a)} \gamma_{GJ}(a)
\f{da}{a\,\beta(a)}}_{=: a_s\tilde{Z}(a_s)} [\partial_\mu J_5^\mu]^{\text{RGI}}\\
=&a_s\left\{ Z_{G\tilde{G}}(a_s) \tilde{O}_1^B+\left(Z_{GJ}(a_s)-E_2(a_s)\tilde{Z}(a_s)Z_J^s(a_s)\right) \partial_\mu J_5^{B\,\mu}\right\}
  \label{O1sRGI}  \end{split}
\ee
fulfilling $\mu^2\f{d}{d\mu^2}[\tilde{O}_1]^{\text{RGI}}=0$. 
In similar way as for \re{dJ5RGI} it can be shown that \re{O1sRGI} is invariant under transformations
\mbox{$[\tilde{O}_1]\rightarrow[\tilde{O}_1]'=Z_1(a_s) [\tilde{O}_1]$.} Even if we allow for redefinitons of the kind
\mbox{$[\tilde{O}_1]\rightarrow[\tilde{O}_1]'=Z_1(a_s) [\tilde{O}_1]+Z_2(a_s) [\partial_\mu J_5^\mu]$} the RGI operator
derived with this method is the same:
\begin{align} 
{[\tilde{O}_1]^{\text{RGV}}}'&=[\tilde{O}_1]^{\text{RGV}}+\f{E_1(a_s)Z_2(a_s)}{E_2(a_s)Z_1(a_s)}
[\partial_\mu J_5^\mu]^{\text{RGI}}\\
\Rightarrow\mu^2\f{d}{d\mu^2}{[\tilde{O}_1]^{\text{RGV}}}'&=\left[\f{E_1(a_s)}{E_2(a_s)} \gamma_{GJ}(a_s)
+\mu^2\f{d}{d\mu^2}\left(\f{E_1(a_s)Z_2(a_s)}{E_2(a_s)Z_1(a_s)}\right)\right][\partial_\mu J_5^\mu]^{\text{RGI}}\\
\Rightarrow{[\tilde{O}_1]^{\text{RGI}}}'&={[\tilde{O}_1]^{\text{RGV}}}'-\underbrace{\left[\left(\int\limits^{a_s(\mu)}\f{E_1(a)}{E_2(a)} \gamma_{GJ}(a)
\f{da}{a\,\beta(a)}\right)+\f{E_1(a_s)Z_2(a_s)}{E_2(a_s)Z_1(a_s)}\right]}_{= a_s\tilde{Z}'(a_s)} [\partial_\mu J_5^\mu]^{\text{RGI}} \notag\\
&=[\tilde{O}_1]^{\text{RGV}}-a_s\tilde{Z}(a_s)[\partial_\mu J_5^\mu]^{\text{RGI}}=[\tilde{O}_1]^{\text{RGI}}{}. 
\end{align}
The leading RGI Wilson coefficient
\be
C^{\text{RGI}}_0(q)=\f{1}{(Q^2)^2}\left.\langle 0|X_t^{\text{RGI}}(q)|0\rangle\right|_{\text{pert}} \ee
in an OPE of the RGI correlator
\be
X_t^{\text{RGI}}(q):= i\int\!\mathrm{d}^4x\,e^{iqx} T\{[\tilde{O}_1]^{\text{RGI}}(x)[\tilde{O}_1]^{\text{RGI}}(0)\}
\label{RGIcorr}
\ee
can now be calculated from the same three bare correlators as $C_0$ and the result for its $Q^2$-derivative is
\be
\begin{split}
Q^2\f{d}{dQ^2}\,C_0^{\text{RGI}}=&
       \f{a_s^2(Q^2) \Ng}{\pi^2}\left[          - 1
        + a_s(Q^2)   \left(
          - \f{97}{12} \ca
          + \f{7}{3} \Nf \tr
          \right) \right. \\ & \left.
       +  \f{a_s(Q^2)}{(11\ca - 4\Nf\tr)}            18 \Nf \tr \cf \right. \\ & \left.
       + a_s^2(Q^2)  \left(
          - \f{51959}{864} \ca^2
          + \f{107}{12} \Nf \tr \cf  
          + \f{3793}{108} \Nf \ca \tr  \right.\right. \\ & \left.\left.
          - \f{251}{54} \Nf^2 \tr^2  
          + \f{55}{8} \zeta_{3} \ca^2
          - 3 \zeta_{3} \Nf \tr \cf
          + \f{1}{2} \zeta_{3} \Nf \ca \tr
          \right)  \right. \\ & \left.
       +  \f{a_s^2(Q^2)}{(11\ca - 4\Nf\tr)}   \left(
           \f{291}{2} \Nf \ca \tr \cf
          - 42 \Nf^2 \tr^2 \cf 
          \right)  \right. \\ & \left.
       + \f{a_s^2(Q^2)}{(11\ca - 4\Nf\tr)^2}   \left(
          - \f{297}{4} \Nf \ca \tr \cf^2
          + \f{475}{4} \Nf \ca^2 \tr \cf \right.\right. \\ & \left.\left.
          - 108 \Nf^2 \tr^2 \cf^2
          - 37 \Nf^2 \ca \tr^2 \cf
          + 4 \Nf^3 \tr^3 \cf
          \right)
\right]{},
\end{split} \label{adlerC0TRGI}
\ee
where the logarithmic pieces have been resummed into $a_s(\mu^2=Q^2)$ for brevity. These terms can easiliy be recovered from the RG equations (see \eqref{reconstructlogs} below).
They have been calculated explicitly however using the above definitions in order to be able to use the RGI condition
\mbox{$\mu^2\f{d}{d\mu^2}\left(Q^2\f{d}{dQ^2}\,C_0^{\text{RGI}}\right)=0$} as a consistency check.

As explained in \cite{Zoller:2012qv} a finite and RGI version of $O_1$ can be defined as
\beq
O_1^{\text{RGI}} := \hat{\beta}(a_s) \, [O_1], \ \ \  \hat{\beta}(a_s) := \frac{-\beta(a_s)}{\beta_0} = 
a_s\left(1+  \sum_{i \ge 1} \frac{\beta_i}{\beta_0} a_s^i \right)
{}. \label{O1RGI}
\eeq
The RGI Wilson coefficient
\be \begin{split}
C_1^{\text{RGI}}(Q^2)=&\f{a_s^2}{\hat{\beta}(a_s)}
\left\{ Z_{G\tilde{G}}^2 C_{1,B}^{(\tilde{O}_1^B, \tilde{O}_1^B)}(Q^2) \right.\\
+&(2  Z_{G\tilde{G}} Z_{GJ}-2  E_2 Z_{G\tilde{G}} Z_J^s \tilde{Z}) C_{1,B}^{(\tilde{O}_1^B, \partial_\nu J_5^{B\,\nu})}(Q^2)\\
+&\left.(Z_{GJ}^2-2 E_2 Z_{GJ} Z_J^s \tilde{Z} 
+(E_2 Z_J^s \tilde{Z})^2)C_{1,B}^{(\partial_\nu J_5^{B\,\nu}, \partial_\nu J_5^{B\,\nu})}(Q^2)\right\}
\end{split} 
\ee
which satisfies
\be C^{\text{RGI}}_1 [O_1]^{\text{RGI}}=C_1 [O_1] \ee in the OPE of \re{RGIcorr}.
The result (again with logarithms resummed into $a_s(\mu^2=Q^2)$) is
\begin{align}
C^{\text{RGI}}_1= & 64 a_s(Q^2) \left\{        1
      + a_s(Q^2)   \left(
           \f{157}{36} \ca
          - \f{5}{9} \Nf \tr
          \right) \right.  \\ & \left.
       +  \f{a_s}{(11\ca-4\Nf\tr)}   \left(
          - \f{17}{2} \ca^2
          - 15 \Nf \tr \cf
          + 5 \Nf \ca \tr
          \right) \right.\notag \\ & \left.
       + a_s^2(Q^2)  \left(
           \f{25945}{1296} \ca^2
          - \f{11}{2} \Nf \tr \cf
          - \f{4355}{648} \Nf \ca \tr
          + \f{25}{81} \Nf^2 \tr^2 \right.\right. \notag \\ & \left.\left. 
          - \f{33}{8} \zeta_{3} \ca^2
          + 3 \zeta_{3} \Nf \tr \cf
          - \f{3}{2} \zeta_{3} \Nf \ca \tr
          \right) \right.\notag \\ & \left.
       + \f{a_s^2}{(11\ca-4\Nf\tr)}   \left(
          - \f{2669}{72} \ca^3
          - \f{785}{12} \Nf \ca \tr \cf
          + \f{955}{36} \Nf \ca^2 \tr \right.\right. \notag \\ & \left.\left. 
          + \f{25}{3} \Nf^2 \tr^2 \cf
          - \f{25}{9} \Nf^2 \ca \tr^2
          \right)
       +  \f{a_s^2}{(11\ca-4\Nf\tr)^2}   \left(
          - \f{10619}{288} \ca^4 \right.\right. \notag \\ & \left.\left.
          + \f{561}{8} \Nf \ca \tr \cf^2
          + \f{1451}{48} \Nf \ca^2 \tr \cf
          + \f{3013}{48} \Nf \ca^3 \tr
          + \f{129}{2} \Nf^2 \tr^2 \cf^2 \right.\right. \notag \\ & \left.\left.
          - \f{301}{6} \Nf^2 \ca \tr^2 \cf
          - \f{211}{8} \Nf^2 \ca^2 \tr^2
          - \f{1}{3} \Nf^3 \tr^3 \cf
          + \f{79}{18} \Nf^3 \ca \tr^3
          \right)  \right.\notag \\ & \left.
       + a_s^3(Q^2)   \left(
           \f{19360399}{186624} \ca^3
          + \f{461}{144} \Nf \tr \cf^2
          - \f{614501}{10368} \Nf \ca \tr \cf \right.\right. \notag \\ & \left.\left.
          - \f{1857805}{31104} \Nf \ca^2 \tr
          + \f{28981}{2592} \Nf^2 \tr^2 \cf 
          + \f{126415}{15552} \Nf^2 \ca \tr^2
          - \f{125}{729} \Nf^3 \tr^3 \right.\right. \notag \\ & \left.\left.
          + \f{55}{8} \zeta_{5} \ca^3
          - 15 \zeta_{5} \Nf \tr \cf^2
          + \f{15}{2} \zeta_{5} \Nf \ca \tr \cf
          + 5 \zeta_{5} \Nf \ca^2 \tr \right.\right. \notag \\ & \left.\left.
          - \f{6893}{144} \zeta_{3} \ca^3
          + \f{145}{12} \zeta_{3} \Nf \tr \cf^2
          + \f{1291}{48} \zeta_{3} \Nf \ca \tr \cf
          - \f{349}{144} \zeta_{3} \Nf \ca^2 \tr\right.\right. \notag \\ & \left.\left.
          - \f{13}{2} \zeta_{3} \Nf^2 \tr^2 \cf
          + \f{121}{36} \zeta_{3} \Nf^2 \ca \tr^2
          \right)
       +  \f{a_s^3(Q^2)}{(11\ca-4\Nf\tr)}   \left(
          - \f{441065}{2592} \ca^4 \right.\right. \notag \\ & \left.\left.
          - \f{109529}{432} \Nf \ca^2 \tr \cf
          + \f{1415}{9} \Nf \ca^3 \tr
          + \f{165}{2} \Nf^2 \tr^2 \cf^2
          + \f{15835}{216} \Nf^2 \ca \tr^2 \cf \right.\right. \notag \\ & \left.\left.
          - \f{7825}{216} \Nf^2 \ca^2 \tr^2 
          - \f{125}{27} \Nf^3 \tr^3 \cf
          + \f{125}{81} \Nf^3 \ca \tr^3
          + \f{561}{16} \zeta_{3} \ca^4
          + \f{291}{8} \zeta_{3} \Nf \ca^2 \tr \cf \right.\right. \notag \\ & \left.\left.
          - \f{63}{8} \zeta_{3} \Nf \ca^3 \tr
          - 45 \zeta_{3} \Nf^2 \tr^2 \cf^2
          + \f{75}{2} \zeta_{3} \Nf^2 \ca \tr^2 \cf
          - \f{15}{2} \zeta_{3} \Nf^2 \ca^2 \tr^2 
          \right) \right. \notag \\ & \left.
       +  \f{a_s^3(Q^2)}{(11\ca-4\Nf\tr)^2}   \left(
          - \f{1667183}{10368} \ca^5
          + \f{29359}{96} \Nf \ca^2 \tr \cf^2 
          + \f{227807}{1728} \Nf \ca^3 \tr \cf \right.\right. \notag \\ & \left.\left.
          + \f{1525313}{5184} \Nf \ca^4 \tr 
          + \f{727}{3} \Nf^2 \ca \tr^2 \cf^2 
          - \f{33923}{144} \Nf^2 \ca^2 \tr^2 \cf
          - \f{129511}{864} \Nf^2 \ca^3 \tr^2 \right.\right. \notag \\ & \left.\left.
          - \f{215}{6} \Nf^3 \tr^3 \cf^2
          + \f{317}{12} \Nf^3 \ca \tr^3 \cf
          + \f{10949}{324} \Nf^3 \ca^2 \tr^3
          + \f{5}{27} \Nf^4 \tr^4 \cf
          - \f{395}{162} \Nf^4 \ca \tr^4 
          \right)  \right. \notag \\ & \left.
       +  \f{a_s^3(Q^2)}{(11\ca-4\Nf\tr)^3}   \left(
          - \f{7623}{16} \Nf \ca^2 \tr \cf^3
          + \f{22121}{32} \Nf \ca^3 \tr \cf^2 
          + \f{31207}{32} \Nf \ca^4 \tr \cf \right.\right. \notag \\ & \left.\left.
          - \f{2079}{4} \Nf^2 \ca \tr^2 \cf^3 
          + \f{13533}{8} \Nf^2 \ca^2 \tr^2 \cf^2
          - \f{29647}{12} \Nf^2 \ca^3 \tr^2 \cf
          - 45 \Nf^3 \tr^3 \cf^3 \right.\right. \notag \\ & \left.\left.
          - \f{1911}{2} \Nf^3 \ca \tr^3 \cf^2
          + 1443 \Nf^3 \ca^2 \tr^3 \cf 
          + 178 \Nf^4 \tr^4 \cf^2
          - 384 \Nf^4 \ca \tr^4 \cf
          + \f{104}{3} \Nf^5 \tr^5 \cf \right.\right. \notag \\ & \left.\left.
          - 2178 \zeta_{3} \Nf^2 \ca^2 \tr^2 \cf^2
          + 2178 \zeta_{3} \Nf^2 \ca^3 \tr^2 \cf
          + 1584 \zeta_{3} \Nf^3 \ca \tr^3 \cf^2
          - 1584 \zeta_{3} \Nf^3 \ca^2 \tr^3 \cf \right.\right. \notag \\ & \left.\left.
          - 288 \zeta_{3} \Nf^4 \tr^4 \cf^2
          + 288 \zeta_{3} \Nf^4 \ca \tr^4 \cf
          \right) \notag
\right\}.
\end{align}
Again an explicit calculation including all logarithmic pieces for an arbitrary scale $\mu$ confirms that indeed \mbox{$\mu^2\f{d}{d\mu^2}\,C_1^{\text{RGI}}=0$}
which is a welcome consistency check.
The full results for the RGI coefficients at a general scale $\mu$ are available at
\mbox{\texttt{\bf http://www-ttp.particle.uni-karlsruhe.de/Progdata/ttp13/ttp13-003/}}.

These full results can now be used to obtain the logarithmic pieces of $Q^2\f{d}{dQ^2}C_0^{\text{RGI}}$ and $C_1^{\text{RGI}}$ at four-loop level.
If a generic RGI quantitiy has the structure
\be \begin{split} Q^{\text{RGI}} &= 
 a_s(\mu)A_1 + a_s^2(\mu)(A_2 + l_{\sss{\mu q}}B_2) + 
  a_s^3(\mu)(A_3 + l_{\sss{\mu q}}B_3 + l_{\sss{\mu q}}^2C_3)\\ &+ 
  a_s^4(\mu)(A_4 + l_{\sss{\mu q}}B_4 + l_{\sss{\mu q}}^2C_4 + l_{\sss{\mu q}}^3D_4)  \\ &+
  a_s^5(\mu)(A_5 + l_{\sss{\mu q}}B_5 + l_{\sss{\mu q}}^2C_5 + l_{\sss{\mu q}}^3D_5 + 
     l_{\sss{\mu q}}^4E_5)+\mathcal{O}(a_s^6) \end{split} \ee
with scale independent coefficients $(A_i, B_i,\dots)$ 
the requirement $\mu^2 \f{d}{d\mu^2} Q^{\text{RGI}}\stackrel{\text{!}}{=}0$ leads to
the conditions
\begin{align}
B_2 &= A_1 \beta_0{}, \notag \\
C_3 &= B_2 \beta_0{},\quad
B_3 = A_1 \beta_1+2 A_2 \beta_0{},\notag \\
D_4 &= C_3 \beta_0{},\quad
C_4 = \f{1}{2}\left(3 B_3 \beta_0+2 B_2 \beta_1\right){},\quad
B_4 = A_1 \beta_2+2 A_2 \beta_1+3 A_3 \beta_0{} \label{reconstructlogs}
\end{align}
which in the cases of $Q^2\f{d}{dQ^2}C_0^{\text{RGI}}$ and $C_1^{\text{RGI}}$ can be used as checks for the result with an arbitrary scale $\mu$
or to reconstruct the logarithmic pieces from the result for $\mu^2=Q^2$. In $\mathcal{O}(a_s^5)$ we find
\begin{align} 
E_5 &= D_4 \beta_0{},\notag\\
D_5 &= \f{1}{3}\left(4 C_4 \beta_0+3 C_3 \beta_1\right){},\notag\\
C_5 &= \f{1}{2}\left(2 B_2 \beta_2+3 B_3 \beta_1+4 B_4 \beta_0\right){},\\
B_5 &= A_1 \beta_3+2 A_2 \beta_2 + 3 A_3 \beta_1 +4 A_4 \beta_0{}.\notag
\end{align}
Using the four-loop $\beta$-function\footnote{The one-loop, two-loop and three-loop results are known from
\cite{Gross:1973id,Politzer:1973fx,Jones:1974mm,Egorian:1978zx,Caswell:1974gg,Tarasov:1980au,Larin:1993tp}.}
of QCD \cite{vanRitbergen:1997va,Czakon:2004bu} the following four-loop contributions 
(for QCD colour factors) are derived:
\be \begin{split}
Q^2\f{d}{dQ^2}C_0^{\text{RGI, 4loop}}&= \f{a_s^5 \Ng}{\pi^2}\left\{
\left(\f{\Nf^3}{54}-\f{11 \Nf^2}{12}+\f{121 \Nf}{8}-\f{1331}{16}  \right) l_{\sss{\mu q}}^3 \right. \\&+ 
\left(\f{7 \Nf^3}{36}-\f{1783 \Nf^2}{144}+\f{21647 \Nf}{96}-\f{19569}{16} \right) l_{\sss{\mu q}}^2  \\&+
\f{1}{(33 - 2 \Nf)^2}\left(
\f{251 \Nf^5}{81}+\f{10 \Nf^4 \zeta_3}{3}-\f{147169 \Nf^4}{432}-330 \Nf^3 \zeta_3 \right. \\&\left.
+\f{108663 \Nf^3}{8}+10890 \Nf^2 \zeta_3-\f{48109321 \Nf^2}{192}-\f{299475 \Nf \zeta_3}{2} \right.\\&\left.\left.
+\f{138470387 \Nf}{64}+\f{5929605 \zeta_3}{8}-\f{450379545}{64}\right) l_{\sss{\mu q}}+\text{const.}\right\},
\end{split}
\ee
\be \begin{split}
C_1^{\text{RGI, 4loop}}&=
64\, a_s^5\left\{
\left( \f{\Nf^4}{1296}-\f{11 \Nf^3}{216}+\f{121 \Nf^2}{96}-\f{1331 \Nf}{96}+\f{14641}{256} \right) l_{\sss{\mu q}}^4 \right.  \\&+ 
\left( \f{5 \Nf^4}{972}-\f{1595 \Nf^3}{2592}+\f{4355 \Nf^2}{216}-\f{293975 \Nf}{1152}+\f{424105}{384} \right) l_{\sss{\mu q}}^3  \\&+ 
\left( \f{25 \Nf^4}{1944}-\f{\Nf^3 \zeta_3}{24}-\f{6937 \Nf^3}{2304}-\f{77 \Nf^2 \zeta_3}{16} +\f{1812625 \Nf^2}{13824}\right.  \\& \left.
+\f{6171 \Nf \zeta_3}{32}-\f{954133 \Nf}{512}-\f{107811 \zeta_3}{64}+\f{12658057}{1536} \right) l_{\sss{\mu q}}^2  \\&+
\f{1}{(33 - 2 \Nf)^2}\left( \f{125 \Nf^6}{2187}-\f{17 \Nf^5 \zeta_3}{18}-\f{457613 \Nf^5}{15552} -\f{4237 \Nf^4 \zeta_3}{54} \right.  \\& \left.
-\f{580 \Nf^4 \zeta_5}{9}+\f{13206877 \Nf^4}{5184}+\f{38583 \Nf^3 \zeta_3}{4} +2695 \Nf^3 \zeta_5 \right.  \\& \left. 
-\f{905734235 \Nf^3}{10368}-\f{1172479 \Nf^2 \zeta_3}{4}-\f{56265 \Nf^2 \zeta_5}{2}+\f{6551159345 \Nf^2}{4608} \right.  \\& \left.
+\f{113749075 \Nf \zeta_3}{32}-\f{459195 \Nf \zeta_5}{4} -\f{16816549087 \Nf}{1536} \right.  \\& \left.\left.
-\f{486694791 \zeta_3}{32}+\f{17788815 \zeta_5}{8}+\f{48864828943}{1536}  \right) l_{\sss{\mu q}}+\text{const.}\right\}.
\end{split}
\ee
For completeness we also give the RGI Wilson coefficients for the correlator
\be
i\int\!\mathrm{d}^4x\,e^{iqx} T\{[\partial_\mu J_5^\mu]^{\text{RGI}}(x)[\partial_\mu J_5^\mu]^{\text{RGI}}(0)\}
=(Q^2)^2 C_0^{JJ, \text{RGI}}+C_1^{JJ, \text{RGI}} [O_1]^{\text{RGI}} +\ldots\,{}.
\label{dJ5corrRGI}
\ee
The results read
\be \begin{split}
Q^2\f{d}{dQ^2}\,C_0^{JJ, \text{RGI}}=& 
       \f{\Ng}{\pi^2} \left[ - \f{a_s^2 }{16} \Nf^2 \tr^2   \right]      
\end{split} \label{adlerC0JJRGI}
\ee
and
\be \begin{split}
C_1^{JJ, \text{RGI}}=4 a_s  \Nf^2 \tr^2 &\left\{
     1 +        a_s   \left(
           \f{157}{36} \ca 
          - \f{5}{9} \Nf \tr
          + \f{11}{12} l_{\sss{\mu q}}  \ca 
          - \f{1}{3} l_{\sss{\mu q}} \Nf \tr
          \right) \right. \\ &\left.
       + \f{a_s}{(11\ca-4\Nf\tr)}   \left(  
  - \f{17}{2} \ca^2
          - 15 \Nf\tr\cf
          + 5 \Nf\ca\tr
\right)
\right\}{}.
\end{split} \label{C1JJRGI}
\ee
The four-loop extension of these results with QCD colour factors are given by
\be \begin{split}
Q^2\f{d}{dQ^2}\,C_0^{JJ, \text{RGI, 4loop}}=& 
       \f{a_s^3 \Ng}{\pi^2} \left[  l_{\sss{\mu q}}\f{\Nf^2(-33 + 2\Nf)}{384}  +\text{const.}\right]      
\end{split} 
\ee
and
\be \begin{split}
C_1^{JJ, \text{RGI, 4loop}}=4 a_s^3 &\left\{
l_{\sss{\mu q}}\f{1}{864}\Nf^2(14166 - 1533\Nf + 20\Nf^2) \right. \\& \left. 
+ l_{\sss{\mu q}}^2\left(\f{121\Nf^2}{64} - \f{11\Nf^3}{48} + \f{\Nf^4}{144}\right)
+\text{const.} \right\}{}.
\end{split}
\ee
\subsection{Numerics} \label{chapNumerics}
We now consider the two cases $\Nf=0$ (pure gluodynamics) and $\Nf=3$ which are most important 
for applications. Furthermore we set $Q^2=\mu^2$, i.e. $l_{\sss{\mu q}}=0$. The numerical results
for $C_1$ and $C_1^{\text{RGI}}$ are then
\begin{align}
C_1(Q^2=\mu^2,\Nf=0)&= 64\{1+13.0833 a_s+135.547 a_s^2+1439.88 a_s^3\},\\
C_1(Q^2=\mu^2,\Nf=3)&= 64\{1+12.25 a_s+94.0971 a_s^2+646.69 a_s^3\},\\
C_1^{\text{RGI}}(Q^2=\mu^2,\Nf=0)&= 64 a_s\{1+10.7652 a_s+102.475 a_s^2+1089.78 a_s^3\},\\
C_1^{\text{RGI}}(Q^2=\mu^2,\Nf=3)&= 64 a_s\{1+9.13889 a_s+55.9532 a_s^2+361.615 a_s^3\}.
\end{align}
In order to estimate the numerical significance of the higher order corrections we evaluate $C_1$ at $\mu=M_Z$,
$\mu=3.5$ GeV and $\mu=2$ GeV with
\be \als^{(\Nf=5)}(M_Z)\approx 0.118 \text{ , } \als^{(\Nf=3)}(3.5 \text{GeV})\approx 0.24
\text{ and } \als^{(\Nf=3)}(2 \text{GeV})\approx 0.30 \text{ \cite{Chetyrkin:2000yt}}\ee
for the cases $\Nf=5$ and $\Nf=3$ respectively.
\begin{align}
C_1(Q^2=\mu^2=M_Z^2,\Nf=5)&= 64\;(
\underbrace{0.0116}_{\text{3 loop}} 
+\underbrace{0.0949}_{\text{2 loop}}  
+\underbrace{0.4393 }_{\text{1 loop}} 
+\underbrace{1}_{\text{0 loop}}),\\
C_1(Q^2=\mu^2=(3.5 \text{ GeV})^2,\Nf=3)&= 64\;(
\underbrace{0.2883}_{\text{3 loop}} 
+\underbrace{0.5492}_{\text{2 loop}} 
+\underbrace{0.9358}_{\text{1 loop}} 
+\underbrace{1}_{\text{0 loop}}),\\
C_1(Q^2=\mu^2=(2 \text{ GeV})^2,\Nf=3)&= 64\;(
\underbrace{0.5631}_{\text{3 loop}} 
+\underbrace{0.8581}_{\text{2 loop}} 
+\underbrace{1.1698}_{\text{1 loop}} 
+\underbrace{1}_{\text{0 loop}}). 
\end{align}
At the scale $\mu^2=M_Z^2$ the two and three-loop contributions are about $9\%$ and $1\%$ wrt tree-level,
whereas at a scale $\mu^2=(2 \text{ GeV})^2$ these contributions become so large that perturbation theory
stops to work (as is expected).
From this evaluation we can assume that in the case of $Q^2=\mu^2$ the Wilson coefficient to this accuracy 
in perturbation theory is a valid approximation down to a scale of about $\mu^2=(3.5 \text{ GeV})^2$.

It is interesting to compare this with the numerics for the Adler function of the coefficent $C_0$, i.e. the purely perturbative part
of the pseudoscalar gluonium correlator:
\begin{align}
\left[Q^2\f{d}{dQ^2}C_0\right](Q^2=\mu^2,\Nf=0)&= -\f{\Ng}{\pi^2}\{1+ 24.25 a_s+466.862 a_s^2 \},\\
\left[Q^2\f{d}{dQ^2}C_0\right](Q^2=\mu^2,\Nf=3)&= -\f{\Ng}{\pi^2}\{1+20.75 a_s+305.953 a_s^2\},\\
\left[Q^2\f{d}{dQ^2}C_0^{\text{RGI}}\right](Q^2=\mu^2,\Nf=0)&= -\f{a_s^2 \Ng}{\pi^2}\{1+24.25 a_s+466.862 a_s^2\},\\
\left[Q^2\f{d}{dQ^2}C_0^{\text{RGI}}\right](Q^2=\mu^2,\Nf=3)&= -\f{a_s^2 \Ng}{\pi^2}\{1+19.4167 a_s+277.194 a_s^2\}.
\end{align}
Evaluated at the same scales as $C_1$ we find:
\begin{align}
\left[Q^2\f{d}{dQ^2}C_0\right](Q^2=\mu^2=M_Z^2,\Nf=5)&= -\f{\Ng}{\pi^2}\;(
\underbrace{0.2967}_{\text{3 loop}} 
+\underbrace{0.6917}_{\text{2 loop}}  
+\underbrace{1}_{\text{1 loop}} 
),\\
\left[Q^2\f{d}{dQ^2}C_0\right](Q^2=\mu^2=(3.5 \text{ GeV})^2,\Nf=3)&= -\f{\Ng}{\pi^2}\;(
\underbrace{1.7856}_{\text{3 loop}} 
+\underbrace{1.5852}_{\text{2 loop}} 
+\underbrace{1}_{\text{1 loop}} 
),\\
\left[Q^2\f{d}{dQ^2}C_0\right](Q^2=\mu^2=(2 \text{ GeV})^2,\Nf=3)&= -\f{\Ng}{\pi^2}\;(
\underbrace{2.7900}_{\text{3 loop}} 
+\underbrace{1.9815}_{\text{2 loop}} 
+\underbrace{1}_{\text{1 loop}} 
). 
\end{align}
We see that the purely perturbative part of the OPE is much less convergent
than $C_1$. In fact it stops to converge already at a scale of about $\mu^2=(20 \text{ GeV})^2$
which corresponds to
\be \als^{(\Nf=4)}(20 \text{GeV})\approx 0.15 \text{ \cite{Chetyrkin:2000yt}}\ee
and hence
\begin{align}
\left[Q^2\f{d}{dQ^2}C_0\right](Q^2=\mu^2=(20 \text{GeV})^2,\Nf=4)&= -\f{\Ng}{\pi^2}\;(
\underbrace{0.5858}_{\text{3 loop}} 
+\underbrace{0.9350}_{\text{2 loop}}  
+\underbrace{1}_{\text{1 loop}} ).
\end{align}
This behaviour should be taken into account in any application that approaches low energies, e.g. in sum rules. We note however that in sum rules, e.g. in \cite{Novikov_scalargluonium,Novikov:1979ux}, 
a Borel transformation is used on the $\f{1}{Q^2}$-series of the OPE. The Borel operator \be \hat{\ssL}_M=\lim\f{(Q^2)^{n}}{(n-1)!}\lb\f{-d}{dQ^2}\rb^n,\; n\rig\infty,Q^2\rig\infty,\f{Q^2}{n}=M^2 \ee strongly
enhances the convergence of the OPE (i.e. the expansion in $\f{1}{Q^2}$) and the scale $\mu$ is then usually set to the finite Borel mass $M$. Nevertheless, the numerical evaluation
presented here suggests that the convergence of the $\als$-expansion is a problem in sum rules using the OPE of the pseudoscalar gluonium correlator at low scales.

\section{Discussion and Conclusions}

I have presented higher order corrections for the coefficient function
$C_1$ of the OPE of the correlator  $X_t$ of two pseudoscalar gluonium operators.
This result extends the previously known accuracy by two loops. It is also
worth of notice that no contact terms can appear in this coefficient due to
the relation between the operator $\tilde{O}_1$ and the Chern-Simons current,
a fact that has been explicitly checked and verified up to $\mathcal{O}(\als^3)$ by this calculation.
The OPE of the correlator of two operators $\partial_\mu J_5^\mu$ which mixes with $\tilde{O}_1$
under renormalization has been performed as well and the corresponding coefficients $C_0^{JJ}$ and $C_1^{JJ}$
have been given at three-loop level.
In addition the construction of RGI operators and Wilson coefficients has been discussed,
the coefficients $C_0^{\text{RGI}}$, $C_1^{\text{RGI}}$, $C_0^{JJ,\text{RGI}}$ and $C_1^{JJ,\text{RGI}}$ have been presented and their
logarithmic part has been derived at four-loop level from the principle of scale invariance.
Finally, the numerical evaluation of $C_0$ and $C_1$ shows large coefficients in the $\als$-expansion
causing a breakdown of the applicability of perturbation theory already at $Q^2=\mu^2=(20 \text{ GeV})^2$ for $Q^2\f{d}{dQ^2}C_0$ and at
$Q^2=\mu^2=(3.5 \text{ GeV})^2$ for $C_1$.

\section*{Acknowledgements}
I am grateful to K.~G.~Chetyrkin for initiating this project, many interesting discussions
and constant support. I also thank J.~H.~K\"uhn for valuable discussions and support.
Finally, I would like to thank M. Jamin, Y. Schr\"oder and A. Vuorinen for useful comments.

All calculations have been
performed on a SGI ALTIX 24-node IB-interconnected cluster of 8-cores
Xeon computers using the thread-based \cite{Tentyukov:2007mu} version  of FORM
\cite{Vermaseren:2000nd}.  The Feynman diagrams  have been drawn with the 
Latex package Axodraw \cite{Vermaseren:1994je}.

This work has been supported by the Deutsche Forschungsgemeinschaft in the
Sonderforschungsbereich/Transregio SFB/TR-9 ``Computational Particle
Physics'' and the Graduiertenkolleg ``Elementarteilchenphysik
bei h\"ochsten Energien und h\"ochster Pr\"azission''

\bibliographystyle{JHEP}

\bibliography{Literatur_v2}

\providecommand{\href}[2]{#2}\begingroup\raggedright\begin{thebibliography}{10}

\bibitem{Chetyrkin:1998mw}
K.~Chetyrkin, B.~A. Kniehl, M.~Steinhauser, and W.~A. Bardeen, {\it {Effective
  QCD interactions of CP odd Higgs bosons at three loops}},  {\em Nucl.Phys.}
  {\bf B535} (1998) 3--18, [\href{http://xxx.lanl.gov/abs/hep-ph/9807241}{{\tt
  hep-ph/9807241}}].

\bibitem{Zoller:2012qv}
M.~Zoller and K.~Chetyrkin, {\it {OPE of the energy-momentum tensor correlator
  in massless QCD}},  \href{http://xxx.lanl.gov/abs/1209.1516}{{\tt
  arXiv:1209.1516}}.

\bibitem{wilson_ope}
K.~G. Wilson, {\it Non-lagrangian models of current algebra},  {\em Phys. Rev.}
  {\bf 179} (1969), no.~5 1499--1512.

\bibitem{Shifman:1978bx}
M.~A. Shifman, A.~I. Vainshtein, and V.~I. Zakharov, {\it {QCD and Resonance
  Physics. Sum Rules}},  {\em Nucl. Phys.} {\bf B147} (1979) 385--447.

\bibitem{forkel_sumrule}
H.~Forkel, {\it Direct instantons, topological charge screening, and qcd
  glueball sum rules},  {\em Phys. Rev. D} {\bf 71} (2005), no.~5 054008.

\bibitem{Luscher:2004fu}
M.~Luscher, {\it {Topological effects in QCD and the problem of short distance
  singularities}},  {\em Phys.Lett.} {\bf B593} (2004) 296--301,
  [\href{http://xxx.lanl.gov/abs/hep-th/0404034}{{\tt hep-th/0404034}}].

\bibitem{PhysRevLett.94.032003}
L.~Del~Debbio, L.~Giusti, and C.~Pica, {\it Topological susceptibility in su(3)
  gauge theory},  {\em Phys. Rev. Lett.} {\bf 94} (Jan, 2005) 032003.

\bibitem{Witten:1979vv}
E.~Witten, {\it {Current Algebra Theorems for the U(1) Goldstone Boson}},  {\em
  Nucl.Phys.} {\bf B156} (1979) 269.

\bibitem{Veneziano:1979ec}
G.~Veneziano, {\it {U(1) Without Instantons}},  {\em Nucl.Phys.} {\bf B159}
  (1979) 213--224.

\bibitem{Seiler:2001je}
E.~Seiler, {\it {Some more remarks on the Witten-Veneziano formula for the
  eta-prime mass}},  {\em Phys.Lett.} {\bf B525} (2002) 355--359,
  [\href{http://xxx.lanl.gov/abs/hep-th/0111125}{{\tt hep-th/0111125}}].

\bibitem{Giusti:2001xh}
L.~Giusti, G.~Rossi, M.~Testa, and G.~Veneziano, {\it {The U(A)(1) problem on
  the lattice with Ginsparg-Wilson fermions}},  {\em Nucl.Phys.} {\bf B628}
  (2002) 234--252, [\href{http://xxx.lanl.gov/abs/hep-lat/0108009}{{\tt
  hep-lat/0108009}}].

\bibitem{Novikov:1979ux}
V.~Novikov, M.~A. Shifman, A.~Vainshtein, and V.~I. Zakharov, {\it {eta-prime
  Meson as Pseudoscalar Gluonium}},  {\em Phys.Lett.} {\bf B86} (1979) 347.

\bibitem{Kataev:1981aw}
A.~Kataev, N.~Krasnikov, and A.~Pivovarov, {\it {The connection between the
  scales of the gluon and quark worlds in perturbative QCD}},  {\em Phys.Lett.}
  {\bf B107} (1981) 115--118.

\bibitem{Baikov:2006ch}
P.~A. Baikov and K.~G. Chetyrkin, {\it {Higgs decay into hadrons to order
  $\alpha_s^5$}},  {\em Phys. Rev. Lett.} {\bf 97} (2006) 061803,
  [\href{http://xxx.lanl.gov/abs/hep-ph/0604194}{{\tt hep-ph/0604194}}].

\bibitem{Kataev:1982gr}
A.~Kataev, N.~Krasnikov, and A.~Pivovarov, {\it {Two loop calculations for the
  propagators of gluonic currents}},  {\em Nucl.Phys.} {\bf B198} (1982)
  508--518, [\href{http://xxx.lanl.gov/abs/hep-ph/9612326}{{\tt
  hep-ph/9612326}}].

\bibitem{Pivovarov_tensorcurrents}
A.~A. Pivovarov, {\it {Two-loop corrections to the correlator of tensor
  currents in gluodynamics}},  {\em Phys. Atom. Nucl.} {\bf 63} (2000)
  1646--1649, [\href{http://xxx.lanl.gov/abs/hep-ph/9905485}{{\tt
  hep-ph/9905485}}].

\bibitem{Laine:2010tc}
M.~Laine, M.~Vepsalainen, and A.~Vuorinen, {\it {Ultraviolet asymptotics of
  scalar and pseudoscalar correlators in hot Yang-Mills theory}},  {\em JHEP}
  {\bf 1010} (2010) 010, [\href{http://xxx.lanl.gov/abs/1008.3263}{{\tt
  arXiv:1008.3263}}].

\bibitem{Novikov_scalargluonium}
V.~A. Novikov, M.~A. Shifman, A.~I. Vainshtein, and V.~I. Zakharov, {\it In
  search of scalar gluonium},  {\em Nuclear Physics B} {\bf 165} (1980), no.~1
  67 -- 79.

\bibitem{Zhang:2003mr}
A.-l. Zhang and T.~G. Steele, {\it {Instanton and higher loop perturbative
  contributions to the QCD sum rule analysis of pseudoscalar gluonium}},  {\em
  Nucl.Phys.} {\bf A728} (2003) 165--181,
  [\href{http://xxx.lanl.gov/abs/hep-ph/0304208}{{\tt hep-ph/0304208}}].

\bibitem{Spiridonov:1984br}
V.~Spiridonov, {\it Anomalous dimension of $g_{\mu\nu}^2$ and
  $\beta$-function},  {\em Preprint IYAI-P-0378} (1984).

\bibitem{Nielsen:1975ph}
N.~Nielsen, {\it {Gauge Invariance and Broken Conformal Symmetry}},  {\em
  Nucl.Phys.} {\bf B97} (1975) 527.

\bibitem{Larin:1993tq}
S.~Larin, {\it {The Renormalization of the axial anomaly in dimensional
  regularization}},  {\em Phys.Lett.} {\bf B303} (1993) 113--118,
  [\href{http://xxx.lanl.gov/abs/hep-ph/9302240}{{\tt hep-ph/9302240}}].

\bibitem{Larin:1993tp}
S.~Larin and J.~Vermaseren, {\it {The Three loop QCD Beta function and
  anomalous dimensions}},  {\em Phys.Lett.} {\bf B303} (1993) 334--336,
  [\href{http://xxx.lanl.gov/abs/hep-ph/9302208}{{\tt hep-ph/9302208}}].

\bibitem{QGRAF}
P.~Nogueira, {\it {Automatic Feynman graph generation}},  {\em J. Comput.
  Phys.} {\bf 105} (1993) 279--289.

\bibitem{Vermaseren:2000nd}
J.~A.~M. Vermaseren, {\it {New features of FORM}},
  \href{http://xxx.lanl.gov/abs/math-ph/0010025}{{\tt math-ph/0010025}}.

\bibitem{Larin:1991fz}
S.~A. Larin, F.~V. Tkachov, and J.~A.~M. Vermaseren, {\it The form version of
  mincer}, . NIKHEF-H-91-18.

\bibitem{MINCER}
S.~G. Gorishnii, S.~A. Larin, L.~R. Surguladze, and F.~V. Tkachov, {\it
  {MINCER: Program for multiloop calculations in quantum field theory for the
  SCHOONSCHIP system}},  {\em Comput. Phys. Commun.} {\bf 55} (1989) 381--408.

\bibitem{vanRitbergen:1998pn}
T.~van Ritbergen, A.~N. Schellekens, and J.~A.~M. Vermaseren, {\it Group theory
  factors for feynman diagrams},  {\em Int. J. Mod. Phys.} {\bf A14} (1999)
  41--96, [\href{http://xxx.lanl.gov/abs/hep-ph/9802376}{{\tt
  hep-ph/9802376}}].

\bibitem{Gorishnii:1983su}
S.~G. Gorishny, S.~A. Larin, and F.~V. Tkachov, {\it {The Algorithm For OPE
  Coefficient Functions In The MS Scheme}},  {\em Phys. Lett.} {\bf B124}
  (1983) 217--220.

\bibitem{Gorishnii:1986gn}
S.~G. Gorishny and S.~A. Larin, {\it {Coefficient Functions Of Asymptotic
  Operator Expansions In Minimal Subtraction Scheme}},  {\em Nucl. Phys.} {\bf
  B283} (1987) 452.

\bibitem{Bos:1992nd}
M.~Bos, {\it {Explicit calculation of the renormalized singlet axial anomaly}},
   {\em Nucl.Phys.} {\bf B404} (1993) 215--244,
  [\href{http://xxx.lanl.gov/abs/hep-ph/9211319}{{\tt hep-ph/9211319}}].

\bibitem{Chetyrkin:1993hk}
K.~G. Chetyrkin and J.~H. K{\"u}hn, {\it {Neutral current in the heavy top
  quark limit and the renormalization of the singlet axial current}},  {\em Z.
  Phys.} {\bf C60} (1993) 497--502.

\bibitem{Gross:1973id}
D.~Gross and F.~Wilczek, {\it {Ultraviolet Behavior of Nonabelian Gauge
  Theories}},  {\em Phys.Rev.Lett.} {\bf 30} (1973) 1343--1346.

\bibitem{Politzer:1973fx}
H.~D. Politzer, {\it {Reliable Perturbative Results for Strong Interactions?}},
   {\em Phys.Rev.Lett.} {\bf 30} (1973) 1346--1349.

\bibitem{Jones:1974mm}
D.~Jones, {\it {Two Loop Diagrams in Yang-Mills Theory}},  {\em Nucl.Phys.}
  {\bf B75} (1974) 531.

\bibitem{Egorian:1978zx}
E.~Egorian and O.~Tarasov, {\it {Two loop renormalization of the QCD in an
  arbitrary gauge}},  {\em Teor.Mat.Fiz.} {\bf 41} (1979) 26--32.

\bibitem{Caswell:1974gg}
W.~E. Caswell, {\it {Asymptotic Behavior of Nonabelian Gauge Theories to Two
  Loop Order}},  {\em Phys.Rev.Lett.} {\bf 33} (1974) 244.

\bibitem{Tarasov:1980au}
O.~Tarasov, A.~Vladimirov, and A.~Y. Zharkov, {\it {The Gell-Mann-Low Function
  of QCD in the Three Loop Approximation}},  {\em Phys.Lett.} {\bf B93} (1980)
  429--432.

\bibitem{vanRitbergen:1997va}
T.~van Ritbergen, J.~Vermaseren, and S.~Larin, {\it {The Four loop beta
  function in quantum chromodynamics}},  {\em Phys.Lett.} {\bf B400} (1997)
  379--384, [\href{http://xxx.lanl.gov/abs/hep-ph/9701390}{{\tt
  hep-ph/9701390}}].

\bibitem{Czakon:2004bu}
M.~Czakon, {\it {The Four-loop QCD beta-function and anomalous dimensions}},
  {\em Nucl.Phys.} {\bf B710} (2005) 485--498,
  [\href{http://xxx.lanl.gov/abs/hep-ph/0411261}{{\tt hep-ph/0411261}}].

\bibitem{Chetyrkin:2000yt}
K.~Chetyrkin, J.~H. Kuhn, and M.~Steinhauser, {\it {RunDec: A Mathematica
  package for running and decoupling of the strong coupling and quark masses}},
   {\em Comput.Phys.Commun.} {\bf 133} (2000) 43--65,
  [\href{http://xxx.lanl.gov/abs/hep-ph/0004189}{{\tt hep-ph/0004189}}].

\bibitem{Tentyukov:2007mu}
M.~Tentyukov and J.~A.~M. Vermaseren, {\it {The multithreaded version of
  FORM}},  \href{http://xxx.lanl.gov/abs/hep-ph/0702279}{{\tt hep-ph/0702279}}.

\bibitem{Vermaseren:1994je}
J.~A.~M. Vermaseren, {\it Axodraw},  {\em Comput. Phys. Commun.} {\bf 83}
  (1994) 45--58.

\end{thebibliography}\endgroup

\end{document}